\documentclass[a4paper]{lipics-v2021}
\nolinenumbers

\usepackage[utf8]{inputenc}
\usepackage[T1]{fontenc} 
\usepackage{amsbsy, amscd, amsfonts, amssymb, amstext, amsmath, mathtools, amsthm}
\usepackage[ruled,linesnumbered]{algorithm2e}
\usepackage{tikz-cd, tikz, tikz-3dplot}
    \usetikzlibrary{arrows,automata,positioning, shapes, shapes.geometric, arrows.meta, calc,decorations.pathreplacing,decorations.markings,arrows.meta, patterns,backgrounds,intersections} 
\usepackage{cite}
    \makeatletter
    \newcommand{\citecomment}[2][]{\citen{#2}#1\citevar}
    \newcommand{\citeone}[1]{\citecomment{#1}}
    \newcommand{\citetwo}[2][]{\citecomment[,~#1]{#2}}
    \newcommand{\citevar}{\@ifnextchar\bgroup{;~\citeone}{\@ifnextchar[{;~\citetwo}{]}}}
    \newcommand{\citefirst}{\@ifnextchar\bgroup{\citeone}{\@ifnextchar[{\citetwo}{]}}}
    
    \makeatother

\newcommand{\be}{\begin{equation}}
\newcommand{\ee}{\end{equation}}
\newcommand{\beq}{\begin{equation*}}
\newcommand{\eeq}{\end{equation*}}
\newcommand{\bt}{\begin{theorem}}
\newcommand{\et}{\end{theorem}}
\newcommand{\bl}{\begin{lemma}}
\newcommand{\el}{\end{lemma}}
\newcommand{\bp}{\begin{proof}}
\newcommand{\ep}{\end{proof}}

\newcommand{\crit}[1][p_i]{\ensuremath{\operatorname{frz}\left(#1\right)}}

\newcommand{\mebr}{\ensuremath{r_{MEB}}}
\newcommand{\permut}{$k$-distance permutation\xspace}

\newcommand{\frozenfiltration}[1][r]{\ensuremath{\tilde{\mathcal{L}}}}

\newcommand{\spread}{\ensuremath{\Phi}}
\newcommand{\maxdim}{\ensuremath{m_{max}}}
\newcommand{\numberofsites}{\ensuremath{\Gamma}}
\newcommand{\mybinom}[2]{\left(\genfrac{}{}{0pt}{}{#1}{#2}\right)}
\everydisplay{\let\binom\mybinom}

\newcommand{\lens}{\operatorname{L}}
\newcommand{\lensfrozen}{\widetilde{\operatorname{L}}}
\newcommand{\kcov}{\operatorname{L}}
\newcommand{\kcovfiltration}{\operatorname{\mathcal{L}}}
\newcommand{\kcovfrozen}{\widetilde{\operatorname{L}}}
\newcommand{\kcovfrozenfiltration}{\widetilde{\operatorname{\mathcal{L}}}}
\newcommand{\cone}{\operatorname{C}}
\newcommand{\infcone}{\cone_{\infty}}
\newcommand{\coneunion}{\operatorname{C}}
\newcommand{\conefiltration}{\operatorname{\mathcal{C}}}
\newcommand{\sparse}{\operatorname{S}}
\newcommand{\sparsefiltration}{\operatorname{\mathcal{S}}}
\newcommand{\discretesparse}{D}
\newcommand{\discretesparsefiltration}{\mathcal{D}}
\newcommand{\maxk}{\mu}

\title{Sparse Higher Order \v{C}ech Filtrations}

\author{Micka\"el Buchet}{Paris, France}{}{}{}
\author{Bianca {B. Dornelas}}{Institute of Geometry, TU Graz, Austria}{bdornelas@tugraz.at}{}{}
\author{Michael Kerber}{Institute of Geometry, TU Graz, Austria}{kerber@tugraz.at}{}{}
\authorrunning{M. Buchet, B.\,B. Dornelas, and M. Kerber}

\Copyright{Micka\"el Buchet, Bianca B. Dornelas, and Michael Kerber}
\funding{This research has been supported by the Austrian Science Fund (FWF), grant
numbers P~33765-N and W1230.}
\acknowledgements{We thank Alexander Rolle for his valuable input and encouragement of this work.}

\begin{document}
\maketitle
\begin{abstract}    
    For a finite set of balls of radius $r$, the $k$-fold cover is the space
    covered by at least $k$ balls. Fixing the ball centers and varying the radius,
    we obtain a nested sequence of spaces that is called the $k$-fold filtration
    of the centers. For $k=1$, the construction is the union-of-balls filtration
    that is popular in topological data analysis. 
    For larger $k$, it yields a cleaner shape reconstruction in the presence of outliers. 
    We contribute a sparsification algorithm  to approximate the topology of the $k$-fold filtration. 
    Our method is a combination and adaptation of several techniques from the well-studied case $k=1$, 
    resulting in a sparsification of linear size that can be computed in expected near-linear time 
    with respect to the number of input points.
    Our method also extends to the multicover bifiltration,
    composed of the $k$-fold filtrations for several values of $k$,
    with the same size and complexity bounds.
\end{abstract}

\section{Introduction}
    \emph{Persistent homology}~\cite{carlsson-topology-and-data, Edels-Harer_Comput-Topo, elz-first-persistent-paper} is a major branch of topological data analysis with applications, for instance, in shape recognition~\cite{oudot_chazal-2016-shape_recog}, material science~\cite{hiraoka-2016-pnas} and biology~\cite{gameiro2015-protein_compressibility,hess_18_bluebrain-neurons}. It studies the homological properties of sequences of topological spaces. A standard construction is to take the homogeneous union of balls, with increasing radius, centered at finitely many points of $\mathbb{R}^d.$  We call these points \emph{sites} and refer to that filtration as the \emph{union-of-balls filtration}. For computational purposes, one considers the homologically equivalent \emph{\v{C}ech filtration}, which is a sequence of simplicial complexes that captures the intersection patterns of the balls in the union-of-balls filtration~\cite[Chap.3]{Edels-Harer_Comput-Topo}\cite{ghrist2014-elementary_applied_topo}.
        
    The drawback of the \v{C}ech filtration (as well as of the closely-related Vietoris-Rips filtration) is that for $n$ sites, it consists of up to $\binom{n}{m+1}$ $m$-simplices because every $(m+1)$-subset of balls intersects at a sufficiently large radius. A technique to overcome this large size is to \emph{approximate} the \v{C}ech (or Vietoris-Rips) filtration with another, much smaller simplicial filtration with similar topological properties. Technically, that means that the persistence modules induced by the homology of the \v{C}ech filtration and its approximation are $\epsilon$-interleaved for an arbitrary $\epsilon>0$~\cite{Chazal-2009-interleavings}. Several strategies have been devised to construct such approximations with total size linear in $n$ for any fixed $\epsilon$ (see related work).
    Many of these approaches work by selecting only a subset of the simplices of the \v{C}ech filtration, in which case we refer to the approximation as a \emph{sparsification}.
        
    The union-of-balls filtration is a special case of the \emph{$k$-fold filtration} built upon the \emph{$k$-fold cover}. For $n$ sites in $\mathbb{R}^d$ and $k\geq 1$ fixed, the $k$-fold cover is the subset of $\mathbb{R}^d$ consisting of points contained in at least $k$ balls of radius $r$ centered at the sites. Besides being a natural extension, $k$-fold filtrations are tightly related with the $k$th neighbor distance that arises in the context of outlier removal and processing of non-homogeneous data densities~\cite{Chazal_Cohen-Steiner_Merigot:11:inference_for_prob,Edelsbrunner_Osang2021-multicover_persistence-rhomboid,Phillips2015-kernel_density,sheehy:12:barycentric_nerve_and_k-fold}. For that reason, they have received increased attention recently, both with regards to computational~\cite{kerber_corbet-2023-multicover_bifiltration,Edels-Osang_Compute-High-Del} and structural aspects~\cite{blumberg-2020-2_parameter_stability}.
    
    For fixed $k$, the $k$-fold filtration can be equivalently expressed by its nerve, which is a simplicial filtration called the \emph{$k$th order \v{C}ech filtration}. It captures the intersection patterns of all $k$-wise intersections of balls, which we call \emph{lenses}. The aforementioned size issue for \v{C}ech filtrations is even more important in the $k$th order case: the filtration is defined over $\binom{n}{k}$ vertices (one for each $k$-subset of sites) and consequently consists of $\mybinom{\binom{n}{k}}{m+1}$ $m$-simplices, making it unrealistic to compute even for small values of $n$. Therefore we need to reduce its size considerably while maintaining a good approximation quality.

    The $k$-fold cover and the higher order \v{C}ech complexes are also studied with relation to \emph{multiparameter persistence}: considering the order $k$ as a second varying parameter, we obtain the \emph{multicover bifiltration}, and the task of approximating the $k$-fold filtration extends to the question of whether there exists a structure that approximates
    the entire multicover bifiltration at once (up to some maximal value $\maxk$). The natural idea of combining sparsified versions of the $k$th order \v{C}ech filtration for different
    values of $k$ bears the technical problem that the set of vertices in order $k$ and $\ell$ are disjoint for $k\neq\ell$. 
    Hence, the approximation scheme must be altered to get compatible approximations for different values of $k$.
        
 \subsection*{Contributions.}
    We propose the first sparsification of the $k$-fold filtration for a fixed $k$. It is a simplicial filtration that, for $n$ sites in $\mathbb{R}^d$ (with constant $d$) and a given parameter $\epsilon>0$, is (multiplicatively) $(1+\epsilon)$-homotopy-interleaved with the $k$-fold filtration. The number of $m$-simplices in our sparsification is 
    \begin{equation}\label{eq-p_simpl_number}\mathcal{O}\left(nk^{k(m+1)}\left(\dfrac{96}{\epsilon}\right)^{\delta k (m+1)}\right),\end{equation}
    where $\delta$ is the doubling dimension of $P$. We point out that for constant $k$ and $\epsilon$, the size of the filtration is linear in the number of sites. This is remarkable because the $k$th order \v{C}ech filtration, which captures the $k$-fold filtration exactly, already contains $\binom{n}{k}$ vertices. Hence our construction avoids including the vast majority of lenses into the sparsification.
    
    We give an output-sensitive algorithm to compute our sparsification up to dimension $\maxdim$ in 
    \begin{equation}\label{eq-complexity_bound}\mathcal{O}\left(nk\log n\log\spread+X k^{k+1}\left(\dfrac{96}{\epsilon}\right)^{k\delta} \cdot (\maxdim+1)\right)\end{equation}
    expected time. Here $\spread$ is the spread of the point set (i.e., the ratio of diameter and smallest distance of two distinct points)
    and $X$ is the size of the output complex, upper bounded by~(\ref{eq-p_simpl_number}) with $m$ replaced by $\maxdim$. 
    Again considering everything but $n$ as constant, we get a running time of $\mathcal{O}(n\log n)$.

    We extend our approach to approximate the multicover bifiltration that contains the $k$-fold filtrations for all $k$ up to some $\maxk$.
    We compute a simplicial bifiltration that is $(1+\epsilon)$-homotopy-interleaved with the multicover, with the same total size
    and the same computational complexity as given in (\ref{eq-p_simpl_number}) and (\ref{eq-complexity_bound}), respectively, 
    replacing $k$ with $\maxk$.
     
 \subsection*{Techniques}
    The seminal work by Sheehy~\cite{Sheehy13-approxim-vietoris} was the first one to introduce a sparsification technique for Vietoris-Rips filtrations yielding linear size
    and $\mathcal{O}(n\log n)$ running time (assuming all other parameters as constant). His technique extends to \v{C}ech complexes as well with minor adaptations.
    Subsequent work~\cite{Spreeman2015-approximate-cech,buchet-sheehy_2016_persist-homol-measures,cavanna-2015-geometric_sparse,dey-2014-simplicial_map,socg-sheehy-2020-sparse_del}
    introduces several extensions, variations, and simplifications of Sheehy's original sparsification; all these works share essentially the same size and complexity bounds.
 
    Our results are achieved by combining several of these techniques used for approximating in the case $k=1$,
    which required non-trivial adaptation for larger values of $k$.
    The main idea is that for every site $p$, we define a \emph{removal radius} such that, for radii larger than this removal radius,
    all lenses involving $p$ are ignored. That means, for larger and larger radii, we construct simplicial complexes with fewer and fewer sites
    to keep the size small. To determine the removal radii of sites, we introduce the \emph{\permut} which is an ordering of the sites based on the distance to the $k$th closest neighbor. The \permut is a generalization of the farthest point sampling~\cite{gonzales_85_greedy-perm} used in some sparsification schemes~\cite{dey-2014-simplicial_map,socg-sheehy-2020-sparse_del} and induces covering and packing properties analogous to those of $\epsilon$-nets.
    
    The idea of removing a lens beyond a certain radius is justified geometrically by the fact that the remaining lenses cover its entire area after a certain radius.
    This is only true, however, if we \emph{freeze} a lens before removing it, that is, keep it unchanged for a short time while the surrounding lenses keep growing.
    This concept was already introduced in~\cite{cavanna-2015-geometric_sparse}, 
    from where we also adapt the elegant technique of lifting the lenses to convex \emph{cones} in $\mathbb{R}^d\times\mathbb{R}$. 
    This lift allows us to reformulate the construction without really removing objects, as removals are not possible in filtrations.

    The extension to multicovers up to $\maxk$ is based on two observations: first, the $\maxk$-distance permutation can be used
    to approximate all $k$-fold covers for $k=1,\ldots,\maxk$ simultaneously, with comparable quality bounds. Second,
    at level $k$, we do not take the nerve only of all $k$-wise intersections, but of all $\ell$-wise intersections with $k\leq\ell\leq\maxk$.
    This makes the simplicial filtrations at different levels compatible; perhaps surprisingly, this enlarged nerve does not cause
    any asymptotic increase in the size bound and the computation time.

    The major geometric predicate for our computation is to determine for a set of growing balls 
    the smallest radius $r$ of intersection.
    The aforementioned freezing of lenses makes this problem
    technically more challenging, as the balls can grow with different speeds.
    The question of how to determine the first intersection in Euclidean space seems to be unaddressed in previous work on sparsifications.
    We provide a solution by dualizing to the computation 
    of the minimal enclosing ball of the ball's centers~\cite{dyer-1992-ball_of_balls,fischer_thesis,Fischer_Gartner-2004-MEB_of_balls,megiddo-1989-ball_of_balls}.
    
  \subsection*{Further related work}

    An alternative line of research by Choudhary et al.~\cite{ckr-improved,ckr-permutahedron,michael_choudhary_sharath-21-approx_rips} defines approximations of \v{C}ech complexes
    which are not sparsifications. They arrive at slightly improved bounds than the sparsification for $k=1$. 
    Approximate filtrations are also actively researched in practice~\cite{blaser_brun-2019-sparse-practice,sparips,tamal_shi_wang-2019-simba,gudhi:RipsComplex,burella-2021-giotto_for_rips}.  
    
    The multicover bifiltration is actively studied as well:
    Blumberg and Lesnick~\cite{blumberg-2020-2_parameter_stability} survey different multiparameter persistence approaches and show a particularly strong stability
    result for multicovers.
    Sheehy~\cite{sheehy:12:barycentric_nerve_and_k-fold} introduces the barycentric bifiltrations, which is equivalent to the multicover but whose size is prohibitively large.
    The question of computing the multicover bifiltration exactly has been studied by Edelsbrunner and Osang~\cite{Edelsbrunner_Osang2021-multicover_persistence-rhomboid}, whose results have been refined by Corbet et al.~\cite{kerber_corbet-2023-multicover_bifiltration}. 
    The latter authors obtain an equivalent bifiltration to the multicover one but has total size (over all choices of $k$) $\mathcal{O}(n^{d+1})$ for $n$ points in $\mathbb{R}^d$~\cite[Prop.~5]{kerber_corbet-2023-multicover_bifiltration}. 
    Their construction rely on using \emph{higher order Voronoi diagrams} and \emph{Delaunay complexes}~\cite{Edels-1986-degree-k-voro}. That reduces size of \v{C}ech complexes, but cannot lead to linear size without further improvements: the Delaunay filtration's $d$-skeleton is of size $\mathcal{O}(n^{\lceil d/2\rceil})$~\cite{Seidel-87-high_del_size}, which is a substantial improvement over the $\mathcal{O}(n^d)$ size of the \v{C}ech $d$-skeleton, but still super-linear for $d\geq 3$.
    Our approximation foregoes those constructions to reduce the size dependency on $n$ further, with the trade-off that we get an exponential dependency on $k$.    
    
  \subsection*{Differences to the conference version}
    The conference version of this paper appeared at the 39th Symposium on Computational Geometry~\cite{bdk-sparse}. 
    Besides another careful revision of the technical contents,
    there are three major differences to that conference version: first, we (slightly) extended the background section
    being more verbose on basic notions of topology to make the paper more accessible for non-experts.
    Second, we integrated the parts of the paper that were left in an appendix due to size limitations.
    Third, the extension to multicovers is a novel result in this journal version.

    \subsection*{Outline.}
    Section~\ref{sec:background} provides background definitions and results. Section~\ref{sec:k-dist_permut} defines a $k$-distance and uses it to construct the \permut of a point set $P$. In Section~\ref{sec:lenses_sparsification} the permutation is used to define a sparse lens filtration that approximates the $k$-fold cover. That results in a nerve filtration that approximates the $k$th order \v{C}ech complex, as shown in Section~\ref{sec:simplicial_sparsification}. The size bound of that filtration is given in Section~\ref{sec:size_analysis}. Section~\ref{sec:algorithm} provides an algorithm for computing the discrete sparse \v{C}ech filtration. The extension to multicovers is discussed in Section~\ref{sec:multicovers}. We conclude with Section~\ref{sec:conclusion}.

    We decided to keep the structure of the conference version in place, that is, discussing the case of fixed $k$ completely and extending to the case of varying $k$ afterwards.
    The reason is that it is conceptually simpler to work with filtrations in one parameter than in two parameters, especially for non-experts in the field.
    This decision comes with the disadvantage that some concepts have to slightly adapted and redefined in Section~\ref{sec:multicovers}.
    
\section{Background}\label{sec:background}
 \subsection*{Lenses and \texorpdfstring{$k$}{k}-fold covers.} 
    Given a point set $P\subseteq\mathbb{R}^d$ and a fixed $k\in\mathbb{N},$ an element $p\in P$ is called a \emph{site} and a \emph{$k$-subset} of $P$ is a subset with $k$ sites. Let $\binom{P}{k}$ be the collection of all $k$-subsets of $P$ and $A\in\binom{P}{k}.$ Let also $B_r(a)$ denote the closed ball centered at $a$ of radius $r$. The \emph{lens} corresponding to the $k$-subset $A$ at scale $r$ is
    \[\lens_r(A):=\underset{a\in A}{\bigcap}B_r(a).\]
    
    The \emph{$k$-fold cover} of $P$ at scale $r$ is the union of lenses at scale $r$ over all $k$-subsets:
    \[\kcov^{(k)}_r:=\underset{A\in\binom{P}{k}}{\bigcup} \lens_r(A).\]
    See Figure~\ref{fig:multicover_example} for an example. 
    We omit $k$ from the notation when it is fixed and write $\kcov_r$ instead.
    
    \begin{figure}
    \centering
    \begin{subfigure}[b]{0.45\textwidth}
        \centering
    \begin{tikzpicture}[scale=1.4]
        \coordinate (a) at (-0.5,-0.2);
        \coordinate (b) at (0.5,0.2);
        \coordinate (c) at (-1,0);
        \coordinate (d) at (-1,0.7);
        \coordinate (e) at (-0.2,1.3);
        \coordinate (f) at (-0.6,-0.7);
        \coordinate (g) at (1,0.5);
        \foreach \sitea/\siteb in {a/b,a/c,a/d,a/e,a/f,a/g,b/c,b/e,b/g,c/d,c/e,c/f,d/e,e/g}{
            \begin{scope}
        		\clip (\sitea) circle (.8);
        		\clip (\siteb) circle (.8);
        		\fill[color=gray!30] (-2,2)rectangle (2,-2);
        	\end{scope}
        }
        \foreach \site in {a,b,c,d,e,f,g}{
            \filldraw (\site) node[inner sep=0cm] (\site) {} circle (1pt);
            \draw[gray!80] (\site) circle (.8);
        }
    \end{tikzpicture}
    \end{subfigure}
    \begin{subfigure}[b]{0.45\textwidth}
        \centering
    \begin{tikzpicture}[scale=1.4]
        \coordinate (a) at (-0.5,-0.2);
        \coordinate (b) at (0.5,0.2);
        \coordinate (c) at (-1,0);
        \coordinate (d) at (-1,0.7);
        \coordinate (e) at (-0.2,1.3);
        \coordinate (f) at (-0.6,-0.7);
        \coordinate (g) at (1,0.5);
        \foreach \sitea/\siteb/\sitec in {a/b/c,a/b/d,a/b/e,a/b/f,a/b/g,a/c/e,a/c/f,a/c/d,a/d/e,b/e/g,c/d/e}{
        	\begin{scope}
        		\clip (\sitea) circle (.8);
        		\clip (\siteb) circle (.8);
        		\clip (\sitec) circle (.8);
        		\fill[color=gray!30] (-2,2)rectangle (2,-2);
        	\end{scope}
    	}
        \foreach \site in {a,b,c,d,e,f,g}{
            \filldraw (\site) node[inner sep=0cm] (\site) {} circle (1pt);
            \draw[gray!80] (\site) circle (.8);
        }
    \end{tikzpicture}
    \end{subfigure}
    \caption{Example of $2$- (left) and $3$-fold (right) covers for a fixed radius.}
    \label{fig:multicover_example}
    \end{figure}

 \subsection*{Filtrations and equivalence.}
    A collection of topological spaces (e.g., subsets of $\mathbb{R}^d$) 
    $\mathcal{X}=\{X_r\}_{r\geq 0}$ is called a \emph{filtration} if for all $r\leq r',$ it holds that $X_r\subseteq{X}_{r'}$.
    In this paper, we will use the caligraphic font throughout to denote filtrations.
    The letter $r$ denotes the \emph{scale parameter} of the filtration.
    For $P$ and $k$ fixed, the previously defined sets $\kcov_r$ for $r\geq 0$ yield a filtration $\kcovfiltration$, the \emph{$k$-fold cover filtration}.
    
    We recall some basic notions from algebraic topology~\cite[Chap.~9]{Munkres-General_Topology}\cite[Chap.~0]{hatcher_algebraic-topo}:
    Two continuous maps $f,g:X\to Y$ between topological spaces are \emph{homotopic}, $f\simeq g$, if there exists a continuous map $H:X\times I\to Y$
    with $H(\cdot,0)=f$ and $H(\cdot,1)=g$ ($H$ is called a \emph{homotopy}). A map $f:X\to Y$ is called a \emph{homotopy equivalence of $X$ and $Y$} if there exists
    a map $g:Y\to X$ such that $f\circ g\simeq \operatorname{id}_Y$ and $g\circ f\simeq \operatorname{id}_X$, where $\operatorname{id}_\cdot$ is the identity function on the corresponding space.

    Let $\mathcal{X}$ and $\mathcal{Y}$ be two filtrations. We say that $\mathcal{X}$ is \emph{(homotopy) equivalent} to $\mathcal{Y}$ if there exists a family of maps $\{f_r\colon X_r \rightarrow Y_r\}_{r\geq 0}$ that are homotopy equivalences of spaces and additionally commute with the 
    inclusion maps of $\mathcal{X}$ and $\mathcal{Y}$. 

\subsection*{Nerves.}
    An \emph{abstract simplicial complex} $K$ over a vertex set $V$ is a collection of non-empty subsets, called \emph{simplices} such that whenever a set $\sigma\subseteq V$
    is in $K$, every non-empty subset of $\sigma$ is also in $K$~\cite[Chap.~3]{Edels-Harer_Comput-Topo}. Note that an abstract simplicial complex can be geometrically
    realized in a sufficiently high-dimensional (Euclidean) space, embedding every abstract simplex as a geometric simplex. Hence, abstract simplicial
    complexes are topological spaces. A simplicial complex $K'\subseteq K$ is called a \emph{subcomplex} of $K$.
    
    For a finite collection $C$ of subsets of $\mathbb{R}^d$, we can define a simplicial complex with vertex set $C$, called the \emph{nerve} of $C$, written as $\operatorname{Nrv}(C)$, 
    as the set of all subsets of $C$
    that have a non-empty mutual intersection. Note that the nerve can contain simplices of larger dimension than $d$.
    As an example, taking $C:=\{\lens_r(A)\mid A\in\binom{P}{k}\}$ for $r$ fixed, we obtain a simplicial complex called the $k$-th order \v{C}ech complex with radius $r$.
    
    The importance of the nerve construction stems from the \emph{Nerve Theorem} (also called \emph{Nerve Lemma}),
    which we state in a simplified form.
    Given a collection $C=\{U^1,\ldots,U^m\}$ of closed and convex subsets of $\mathbb{R}^d$, we denote by $\bigcup C$ the union
    of all $U^i$. Then, there is a homotopy equivalence between $\bigcup C$ and $\operatorname{Nrv}(C)$.
    Informally, this means that the nerve of $C$ contains the same topological information as the union of the sets.
    This is useful, because the nerve is a combinatorial object (an abstract simplicial complex) that is more viable
    for computation than a subset of $\mathbb{R}^d$.
    
    The Nerve theorem extends to filtrations, in the form of the Persistent Nerve Theorem~\cite[Thm.~3.9]{alex_michael-2020-closed_nerve_thm}.
    Here, we consider a finite collection of filtrations $\{U^{i}_r\}_{r\geq 0}$ over $\mathbb{R}^d$ with $i$ ranging from $1$ to $m$.
    On the one hand, defining $U_r:=\bigcup_{i=1}^m U^{i}_r$, we obtain a filtration of subsets of $\mathbb{R}^d$, called the \emph{union filtration}.
    On the other hand, writing $N_r:=\operatorname{Nrv}\{U^{i}_r\mid i=1,\ldots,m\}$ yields a filtration of simplicial complexes,
    called the \emph{nerve filtration}. The Persistent Nerve Lemma states that if $U^i_r$ is closed and convex for every choice of $i$ and $r$,
    then the union filtration and the nerve filtration are equivalent.
    
    As an example, we pick the lenses $\{\lens_r(A)\}_{r\geq 0}$ with $A$ ranging over all $k$-subsets of $P$. This yields closed and convex
    filtrations, the union filtration is the $k$-fold cover $\kcovfiltration$, and the nerve filtration is the collection of \v{C}ech complexes,
    called the $k$-th order \v{C}ech filtration. The Persistent Nerve Lemma states that these filtrations are equivalent, so the \v{C}ech filtration
    is a combinatorial description of the $k$-fold cover that we are interested in. Unfortunately, the \v{C}ech filtration is too large
    for practical purposes.

 \subsection*{Interleavings.}
    Let $\epsilon\geq 0$. 
      Two filtrations $\mathcal{X}$ and $\mathcal{Y}$ are \emph{(multiplicatively) $(1+\epsilon)$-interleaved} if there exist maps $\{f_{r}:X_r\to Y_{(1+\epsilon)r},g_{r}:Y_r\to X_{(1+\epsilon)r}\}_{r\geq 0}$ such that the diagram
    \[\begin{tikzcd}
        X_r \arrow[rr,hook] \arrow[rd, rightarrow,"f_r" {description, anchor=south,rotate=0,yshift=0.2ex,xshift=1ex}]&       & {X}_{r(1+\epsilon)^2}  \arrow[rd, rightarrow,"f_{r(1+\epsilon)^2}" {description, anchor=south,rotate=0,yshift=0.2ex,xshift=3.5ex}] \arrow[rr,hook] & & {X}_{r(1+\epsilon)^4}\\
         & {Y}_{r(1+\epsilon)} \arrow[rr,hook] \arrow[ru, rightarrow,"g_{r(1+\epsilon)}" {description, anchor=south,rotate=0,yshift=-2ex,xshift=5ex}]&   & {Y}_{r(1+\epsilon)^3} \arrow[ru, rightarrow,"g_{r(1+\epsilon)^3}" {description, anchor=south,rotate=0,yshift=-2ex,xshift=5ex}]  
    \end{tikzcd}\]
    commutes for all $r$. 
    Informally, interleaved filtrations with small $\epsilon$ are good approximations of each other.
    An important special case that occurs in this paper is when $X_r\subset Y_{r}\subseteq X_{r(1+\epsilon)}$;
    in this case $\mathcal{X}$ and $\mathcal{Y}$ are $(1+\epsilon)$-interleaved, choosing the inclusion maps as $f$ and $g$.
    If $\mathcal{X}$, $\mathcal{Y}$ are $(1+\epsilon_1)$-interleaved and $\mathcal{Y}$, $\mathcal{Z}$ are  $(1+\epsilon_2)$-interleaved,
    then $\mathcal{X}$ and $\mathcal{Z}$ are   $(1+\epsilon_1)(1+\epsilon_2)$-interleaved.

    We use a derived, more relaxed notion of approximation~\cite[Def 3.5]{blumberg2022universality}:
    Two filtrations $\mathcal{X}$ and $\mathcal{Y}$ are \emph{(multiplicatively) $(1+\epsilon)$-homotopy-interleaved} if there exist filtrations $\mathcal{X}'$ and $\mathcal{Y}'$ such that 
    $\mathcal{X}'$ is equivalent to $\mathcal{X}$,
    $\mathcal{Y}'$ is  equivalent to $\mathcal{Y}$ and
    $\mathcal{X}'$ and $\mathcal{Y}'$ are $(1+\epsilon)$-interleaved.
    In other words, $\mathcal{X}$ and $\mathcal{Y}$ are interleaved up to homotopy equivalence. 
    Note that two filtrations are $1$-homotopy-interleaved if and only if they are equivalent.
    It is non-trivial, but true, that
    if $\mathcal{X}$ and $\mathcal{Y}$ are $(1+\epsilon_1)$-homotopy-interleaved and $\mathcal{Y}$ and $\mathcal{Z}$ are $(1+\epsilon_2)$-homotopy-interleaved,
    then $\mathcal{X}$ and $\mathcal{Z}$ are $(1+\epsilon_1)(1+\epsilon_2)$-homotopy-interleaved~-- see~\cite[Sec 4]{blumberg2022universality} for a proof.
    
    In this paper, we use the notion ``$\mathcal{X}$ is $(1+\epsilon)$-homotopy-interleaved with $\mathcal{Y}$'' 
    as synonym for ``$\mathcal{X}$ is an $(1+\epsilon)$-approximation of $\mathcal{Y}$''; we will still use the homotopy-interleaving notation
    in the technical parts for concreteness. We remark that being $(1+\epsilon)$-homotopy-interleaved implies that the induced 
    persistence modules (see~\cite[Chap.~7]{Edels-Harer_Comput-Topo}) are (multiplicatively) $(1+\epsilon)$-interleaved
    in the sense of~\cite{Chazal-2009-interleavings}. 
    Persistence modules are a major concept in the field of topological data analysis,
    and this notion of approximation used in most previous papers
    on approximating \v{C}ech and Vietoris-Rips filtrations.

 \subsection*{Geometric notions.}
The \emph{doubling constant} $\Delta$ of a metric space is the minimal integer $\Delta$ such that any ball of arbitrary radius $r$ can be covered with at most $\Delta$ balls of radius $r/2$. The \emph{doubling dimension} of the metric space is $\delta:=\log_2 \Delta$. The Euclidean space $\mathbb{R}^d$
has a doubling dimension of order $\Theta(d)$ and hence constant in this paper. 
However, interpreting a finite point set $P$ in $\mathbb{R}^d$ as a finite metric space, its doubling dimension can be significantly smaller than $d$, for instance if the points all lie close to a low-dimensional subspace.

    To cover a ball $B$ of radius $r$ with balls of radius $r/4,$ one needs at most $\Delta^2$ balls; with balls of radius $r/8$ one needs $\Delta^3$ balls and so on. Thus, to cover $B$ with balls of radius $r',$ we have to find the smallest $t$ such that $r/2^t \leq r'$. That is $t=\lceil \log_2 r/r'\rceil.$ 
    Then,
    $\Delta^t\leq \Delta^{\log_2 r/r' +1} = 2^{\delta} \left({r}/{r'}\right)^{\delta}$
    and $(2r/r')^{\delta}$ balls of radius $r'$ are sufficient to cover $B$.

    The \emph{spread} of a point set $P$, denoted by $\spread$, is the ratio of the diameter of the point set to the shortest distance between two
    distinct points in the point set. Note that the spread can be arbitrarily large already for three points and is therefore
    a quantity that is independent of the cardinality of $P$. The spread appears often in the analysis of geometric approximation
    algorithms that are based on hierarchical grids (such as quadtrees)  because it typically determines the number
    of resolutions that need to be considered by the approximation algorithm~-- see, for instance,~\cite{Har-Peled_book_geometric-approx} for details.

\subsection*{Quadtreaps.}
    A \emph{quadtreap}~\cite{mount_park-2010socg-dynamic_approx_range} is a dynamic data structure for spherical range search in Euclidean space. We summarize its
    properties in a simplified form suitable for us: for a set $X$ of $n$ points in $\mathbb{R}^d$ (with $d$ constant), it can be built in $\mathcal{O}(n\log n)$ expected time. 
    It supports insertions and deletions of points in $X$ in expected $\mathcal{O}(\log n)$ time.
    Moreover, given a query point $q$ and a radius $r$, it returns a list $S\subseteq X$ which is guaranteed to contain all
    points in $X$ of distance $\leq r$ from $q$, and is guaranteed not to contain any point in $X$ of distance $\geq 2r$
    from $q$. The running time for such a query is $\mathcal{O}(\log n+|S|)$.

\section{\texorpdfstring{\permut}{k-distance}}\label{sec:k-dist_permut} 
    Given some integer $k\geq 1$ and a finite data set $P\subseteq\mathbb{R}^d$ 
    of $n\geq k$ sites, we  
    define an order on the points in $P$ in which the sites
    are denoted by $p_1,\ldots,p_n$. Writing $P_i:=\{p_1,\ldots,p_i\}$,
    our order ensures that the $k$-fold cover over $P_i$ approximates
    the $k$-fold cover over $P$, with increasing approximation quality
    when $i$ increases.
    
    The \emph{$k$-distance} of $x\in\mathbb{R}^d$ to $P$, denoted by $\operatorname{d}^k(x,P)$, 
    is the distance from $x$ to its $k$th closest neighbor in $P$.
    We define the \emph{\permut} incrementally as follows:
    we choose $p_1,\ldots,p_k$ as arbitrary, pairwise distinct sites from $P$.
    If $p_1,\ldots,p_{i-1}$ are chosen for $k< i\leq n$, we set
    \[
    p_{i}:=\underset{q\in P\setminus P_{i-1}}{\operatorname{argmax}}\operatorname{d}^k(q,P_{i-1}).
    \]
    Note that for $k=1$, we obtain the well-known farthest point sampling.
    We also define 
    \[
    \lambda_i:=\operatorname{d}^k(p_i,P_{i-1})
    \]
    for $k+1\leq i\leq n$ and set $\lambda_1,\ldots,\lambda_k$ to $\infty$, so that the sequence $(\lambda_{1},\lambda_{2},\cdots,\lambda_{n})$ is non-increasing.
    The next two properties of the \permut are reminiscent of the packing and covering properties of $\epsilon$-nets~\cite[Chap.~14]{sutherland_book_metric-and-topo-spaces}.

    \begin{lemma}[Covering]\label{lem_covering}
        Let $\kcov_r|_{P_i}$ be the $k$-fold cover of $P_i$. Then, for all $k\leq i\leq n-1$, 
        \[\kcov_r|_{P_i}\subseteq \kcov_r \subseteq \kcov_{r+\lambda_{i+1}}|_{P_i}.\]
    \end{lemma}
    
    \begin{proof}
        $P_i\subseteq P$ immediately implies $\kcov_r|_{P_i}\subseteq\kcov_r.$ Consider $x\in\kcov_r.$ Then, $x\in \lens_r(A)$ for some $A=\{a_1,a_2,\ldots, a_k\}\subseteq P$. If $A\subseteq P_i$, the result follows. Otherwise, without loss of generality let $a_1\notin P_i$. By definition of $\lambda_{i+1}$, $\operatorname{d}^k(a_1,P_{i})\leq \lambda_{i+1}$ and hence there are sites $b_1,b_2,\ldots, b_k\in P_i$ with $\operatorname{d}(a_1,b_j)\leq\lambda_{i+1}$ for all $1\leq j\leq k$. Consequently, $\operatorname{d}(x,b_j)\leq\operatorname{d}(x,a_1)+\operatorname{d}(a_1,b_j)\leq r +\lambda_{i+1}$
        and the $k$ closest sites to $x$ in $P_i$ are within distance $r+\lambda_{i+1}$ of $x$, implying $x\in \kcov_{r+\lambda_{i+1}}|_{P_i}$.
    \end{proof} 
    
    \begin{lemma}[Packing]\label{lem_packing}
        For all $k+1\leq i \leq n$, each $p\in P_i$ has
        $\operatorname{d}^k(p,P_i\setminus\{p\})\geq {\lambda_{i}}/{2}$.  
    \end{lemma}
    
    \begin{proof}
        We do induction on $i$. For $i=k+1$, let $q$ be the $k$th closest neighbor of $p_{k+1}$ in $P_k$. We have
        $\operatorname{d}^k(p_{k+1},P_{k+1}\setminus\{p_{k+1}\})=\lambda_{k+1}\geq \lambda_{k+1}/2$
        and, for any $p\in P_{k+1}\setminus\{p_{k+1}\}$,
        \[\operatorname{d}^k(p,P_{k+1}\setminus\{p\})=\max\limits_{p'\in P_{k+1}\setminus\{p\}} \operatorname{d}(p,p')\geq \dfrac{\operatorname{d}(p,q)+\operatorname{d}(p,p_{k+1})}{2} \geq \dfrac{\operatorname{d}(q,p_{k+1})}{2}=\dfrac{\lambda_{k+1}}{2}. \]
        
        Hence the statement is true for $i=k+1$. Next we assume, for some $i\geq k+1$, that for every $p\in P_{i},$ $\operatorname{d}^k(p,P_{i}\setminus\{p\})\geq \lambda_{i}/2$, and show the statement for $i+1$.
    
        For $p_{i+1}$, we have $\operatorname{d}^k(p_{i+1},P_{i+1}\setminus\{p_{i+1}\})=\lambda_{i+1}\geq \lambda_{i+1}/2$ and the statement follows.
        Consider $p\in P_{i+1}\setminus\{p_{i+1}\}$. If $p_{i+1}$ is not among the $k$ nearest neighbors of $p$ in $P_{i+1}$, then
        \[\operatorname{d}^k(p,P_{i+1}\setminus\{p\})=\operatorname{d}^k(p,P_i\setminus\{p\})\geq\dfrac{\lambda_{i}}{2}\geq\dfrac{\lambda_{i+1}}{2}\]
        by the induction hypothesis and because the $\lambda$-values are non-increasing.
        Otherwise, $p_{i+1}$ is among the $k$ nearest neighbors of $p$ in $P_{i+1}\setminus\{p\}$ and
        $\operatorname{d}^k(p, P_{i+1}\setminus\{p\})\geq \operatorname{d}(p,p_{i+1}).$
    
        If $\operatorname{d}(p,p_{i+1})\geq \lambda_{i+1}/2$, the claim follows. Otherwise, every site at distance smaller than $\lambda_{i+1}/2$ of $p$ is at distance smaller than $\lambda_{i+1}$ of $p_{i+1}$. Since $\lambda_{i+1}=\operatorname{d}^k(p_{i+1},P_i)$, there can be at most $k-2$ sites of $P_i\setminus\{p\}$ at distance smaller than $\lambda_{i+1}$ of $p_{i+1}$.
        Thus, counting $p_{i+1}$ as well, there can be at most $k-1$ sites of $P_{i+1}\setminus\{p\}$ at distance smaller than $\lambda_{i+1}/2$ of $p$ and it follows that $\operatorname{d}^k(p,P_{i+1}\setminus\{p\})\geq \lambda_{i+1}/2$.
    \end{proof}

 \subsection*{Computation.}
    We give an algorithm for computing the \permut that is inspired by the algorithm given by Har-Peled and Mendel~\cite[Sec.~3.1]{Har-Peled_2006_nets-algorithm} 
    for the case $k=1$. We call a site \emph{ordered} if it has already been assigned its index in the \permut and \emph{unordered} otherwise. 
        
    For initialization, pick $k$ sites $p_1,\ldots,p_k$ arbitrarily  and compute, for each $y\in P\setminus P_k$, the distances from $y$ to $p_i$, $1\leq i\leq k$. 
    For each $y$, we store these distances in a max-heap~\cite[Chap. 6]{Algorithms-Intro-Book_2009} $T_y$ that also has a fixed label identifying $y$. 
    Store these elements $T_y$ in another max-heap $M$, whose priority value for each $T_y$ is the maximal value in the heap $T_y$.
    Also, initialize a quadtreap $Q$ (see Section~\ref{sec:background}) with the $(n-k)$-unordered sites.
    Up until this point we need $\mathcal{O}(n (\log n+k\log k))$ time. 
    
    The algorithm will proceed in iterations, choosing the next $p_i$ to be ordered in every iteration.
    It will maintain the invariant that, at the beginning of an iteration, $M$ and $Q$ contain exactly one element for each unordered site $y$,
    and the corresponding max-heap $T_y$ in $M$ contains the distances to the $k$ closest ordered sites from $y$.
    
    To describe an iteration, pick the maximal $T_y$ in $M$ and set $p_i:=y$. Let $\lambda_i$ be the maximal value in $T_y$. Remove $T_y$ from $M$ and $y$ from $Q$. 
    Then, query $Q$ for $p_i$ and radius $\lambda_i$, obtaining a list $S_i$ of sites that contains all sites that are at most $\lambda_i$ away from $p_i$, and only contains sites
    that are at most $2\lambda_i$ away from $p_i$.
    For each site $x\in S_i$, add the distance of $x$ to $p_i$ in $T_x$. Then, remove the maximal element from $T_x$. This ends the description of the algorithm.
    
    For correctness, the invariants imply that the $p_i$ as chosen is indeed the unordered site with the largest $k$-distance to the set of ordered sites,
    as required. It is clear that $M$ and $Q$ are correctly updated, and for every site $x$ in $S_i$, the heap $T_x$ is correctly updated.
    It remains to argue that for a site $x\notin S_i$, the addition of $p_i$ to the ordered set does not affect $T_x$.
    But that follows from the fact that for such an $x$, we have $\operatorname{d}^k(x,P_{i-1})\leq\lambda_i$,
    so there are $k$ sites in $P_{i-1}$ in distance $\lambda_i$ or closer to $x$. Since $d(x,p_i)\geq\lambda_i$,
    the heap $T_x$ still contains the $k$ closest distances in $P_i$.
    
    We proceed to the complexity analysis. In iteration $i$, we have a complexity of $\mathcal{O}(\log n)$ to update $M$ and $Q$, $\mathcal{O}(\log n+|S_i|)$
    to generate $S_i$ and $\mathcal{O}(|S_i|\log n)$ to update the heaps $T_s$ (which requires an update of $M$, too). Clearly, the last
    complexity dominates the others, and we obtain a complexity of $\mathcal{O}(|S_i|\log n)$ per iteration.
    Writing $\Sigma$ for sum of all $|S_i|$ in the algorithm, 
    the expected running time is bounded by $\mathcal{O}(n k\log k + n\log n + \Sigma\log n)$ (with a constant that depends exponentially on
    the doubling dimension of $P$). It remains to bound $\Sigma$. Recall that $\spread$ is the spread of $P$.
    
    \begin{lemma}
        Every site appears in at most $\mathcal{O}(k\log\spread)$ sets $S_i$.
    \end{lemma}
    \begin{proof}
        Consider the interval $(2^s,2^{s+1}]$ for some integer $s$ and the sites $p_i,\ldots,p_j$ with $i\leq j$
        with $\lambda_i,\ldots,\lambda_j \in (2^s,2^{s+1}]$. Note that there are at most $\log\spread$  intervals of that form
        with at least one $\lambda$-value inside. Thus it is enough to show that each site appears in at most $\mathcal{O}(k)$ sets among $S_i,\ldots,S_j$.

        Let $p$ be an arbitrary site and fix $i\leq \ell\leq j$, meaning that $\lambda_{\ell}\in (2^s,2^{s+1}]$. To be in the list $S_\ell$, $p$ must be at distance at most $2\cdot2^{s+1}$ from $p_\ell$.
        Since this holds for all $i\leq \ell\leq j$, all sites among $p_i,\ldots,p_{j}$ for which $p$ is reported lie in a ball $B$ of radius $2^{s+2}$ (centered at $p$).
        The (discrete) ball $B\cap P$ can be covered by $32^\delta=\mathcal{O}(1)$ (discrete) balls of radius $2^{s-2}$ (see \emph{Doubling Dimension} in Section~\ref{sec:background}). We show next that every such ball of radius $2^{s-2}$ contains at most $k$ sites of $p_i,\ldots,p_j$, implying the desired bound.
        Let $B'$ be one the balls of radius $2^{s-2}$ covering $B$ and $p_\ell$ be inside $B'$. If the $k$ closest sites of $p_{\ell}$ in $P_j$ are also in $B'$, we have that
        $\operatorname{d}^k(p_\ell,P_j\setminus\{p_\ell\})\leq 2\cdot 2^{s-2}=\dfrac{2^s}{2}<\dfrac{\lambda_j}{2}$, which is a contradiction to the Packing property (Lemma~\ref{lem_packing}). 
    \end{proof}

    The lemma yields the bound $\Sigma=\mathcal{O}(nk\log\spread)$, which implies our final complexity bound.
    
    \begin{theorem}\label{thm:runtime_k_greedy}
        The \permut can be computed in expected time
        $
        \mathcal{O}(nk\log n\log\spread),
        $
        with $\spread$ the spread of the point set.
    \end{theorem}

\section{A sparse union of lenses}\label{sec:lenses_sparsification}
\begin{figure}[ht!]
        \centering
        \includegraphics[scale=0.9]{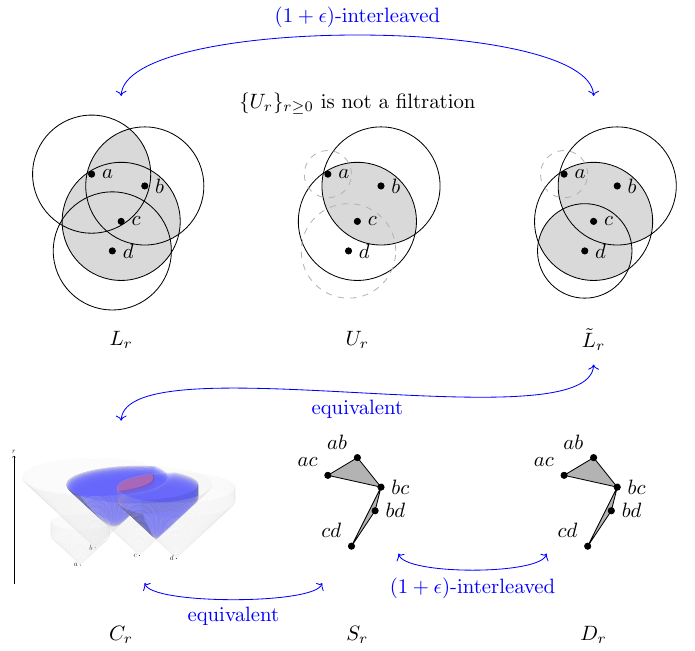}
        \caption{Schematical view of the different filtrations introduced in Sections~\ref{sec:lenses_sparsification} and~\ref{sec:simplicial_sparsification}. We consider $k=2$, $r>(1+\epsilon)\crit[a]$ and $r\in(\crit[d],(1+\epsilon)\crit[d]]$.}
        \label{fig:schematics}
    \end{figure}

    In the following two sections, we will 
    define several filtrations to arrive at our
    final approximation.
    Figure~\ref{fig:schematics} shows a schematic overview of our construction.
    
    Recall that the $k$-fold cover is defined as the union of all lenses at radius $r$, where every lens is given by $k$ sites.
    For large values of $r$, most of these lenses intersect, yielding a size explosion in their nerve.
    At the same time, many lenses are eventually covered by the union of other lenses and so may be removed from consideration.
    
    To define the precise threshold for removal of a lens, recall that in Section~\ref{sec:k-dist_permut} 
    we ordered the sites as $p_1,\ldots,p_n$ and obtained values $\lambda_1,\lambda_2,\ldots,\lambda_k=\infty,\,\lambda_{k+1}\geq\ldots\geq\lambda_n$.
    Fix $\epsilon\in(0,1]$. 
    The \emph{freezing radius} of a site $p_i$ is
    \[\crit[p_i]:=\dfrac{(1+\epsilon)\lambda_{i}}{\epsilon}.\]
    We extend the definition to lenses by setting
    $\crit[A]=\min\limits_{p\in A}\crit[p].$
    Then at radius $r$ we only consider lenses whose freezing radius is at least $r$ and set 
    \[U_r:=\bigcup_{\crit[A]\geq r} \lens_r(A).\] 
    Notice that $\{U_r\}_{r\geq 0}$  is \emph{not} a filtration: Figure~\ref{fig_counterexample_freezing}
    illustrates that $U_r$ might not be a subset of $U_{r'}$ for $r<r'$. Even so, 
    we show a useful result to relate $\kcov_r$ and $U_r$.

    \begin{figure}[ht!]
        \centering
        \begin{subfigure}[b]{0.45\textwidth}
        \centering
        \begin{tikzpicture}[scale=0.8]
            \coordinate (a) at (1.4,-0.3);
            \coordinate (b) at (0.4,-0.8);
            \coordinate (c) at (1,0);
            \begin{scope}
                    \clip (c) circle (2.25cm);
                    \foreach \site in {a,b} 
                        \fill[gray!30] (\site) circle (2.25cm);
            \end{scope}
            \begin{scope}
                    \clip (b) circle (2.25cm);
                    \foreach \site in {a} 
                        \fill[gray!30] (\site) circle (2.25cm);
            \end{scope}
            \filldraw[gray] (c) node[inner sep=0cm,label={[label distance=0.7pt]{115}:{$c$}}] (c) {} circle (1.5pt);
            \foreach \site in {b,a} 
                \filldraw (\site) node[inner sep=0cm,label={[label distance=0.8pt]{0}:{$\site$}}] (\site) {} circle (1.5pt);
            \foreach \site in {a,b,c} 
                \draw (\site) circle (2.25cm);
        \end{tikzpicture}
        \end{subfigure}
        \begin{subfigure}[b]{0.45\textwidth}
        \centering
        \begin{tikzpicture}[scale=0.8]
            \coordinate (a) at (1.4,-0.3);
            \coordinate (b) at (0.4,-0.8);
            \coordinate (c) at (1,0);           
            \begin{scope}
                    \clip (b) circle (2.26cm);
                    \foreach \site in {a} 
                        \fill[gray!30] (\site) circle (2.26cm);
            \end{scope}
            \filldraw[gray] (c) node[inner sep=0cm,label={[label distance=0.7pt]{115}:{$c$}}] (c) {} circle (1.5pt);
            \foreach \site in {a,b} 
                \filldraw (\site) node[inner sep=0cm,label={[label distance=0.8pt]{0}:{$\site$}}] (\site) {} circle (1.5pt);
            \draw[gray!30] (c) circle (2.26cm);
            \foreach \site in {a,b} 
                \draw (\site) circle (2.26cm);
        \end{tikzpicture}
        \end{subfigure}
        \caption{Example in $\mathbb{R}^2$ with $k=2$. \textit{Left:} $U_r$ at radius $r=\crit[c].$  \textit{Right:} $U_{r'}$ at radius immediately after $\crit[c]$. Even though $r<r',$ $U_r\nsubseteq U_{r'}.$}
        \label{fig_counterexample_freezing}
    \end{figure}
    
    \begin{lemma}\label{sandwich_first}
    $U_r\subseteq \lens_r\subseteq U_{(1+\epsilon)r}.$
    \end{lemma}
    
    \begin{proof}        
        The first inclusion is clear. For the second inclusion, consider $x\in\kcov_r$ and let $i$ be the maximal index such that
        \[r\leq \frac{\lambda_{i}}{\epsilon}.\tag{$\ast$}\]
        If $i=n$, then there is $A\subseteq P_n$ with $x\in\lens_r(A)$ because $x\in\kcov_r$ and $P=P_n$. By definition of the freezing radius and
        inequality $(\ast)$,
        $\crit[A]\geq\crit[p_i]= (1+\epsilon)\lambda_i /\epsilon \geq r (1+\epsilon)$ and thus $\lens_{r(1+\epsilon)}(A)\subseteq U_{(1+\epsilon)r}$. Since $\lens_r(A)\subseteq \lens_{(1+\epsilon)r}(A)$, the result follows.
        
        For $i<n$, notice that the Covering Property (Lemma~\ref{lem_covering}) guarantees that
        $x$ is contained in a lens $\lens_{r+\lambda_{i+1}}(A)$ for some $A\subseteq P_i$.
        Since $i$ is maximal, $\lambda_{i+1} / \epsilon< r$ and so 
        $\lens_{r+\lambda_{i+1}}(A)\subseteq\lens_{(1+\epsilon)r}(A)$.
        Moreover, $A\subseteq P_i$ and inequality $(\ast)$ imply $\crit[A]\geq\crit[p_i]\geq(1+\epsilon)r$.
        Hence the lens of $A$ contributes to $U_{(1+\epsilon)r}$
        and as it contains $x$, the statement follows.
    \end{proof}
    
    We next adjust the definition of $U_r$ to obtain an actual filtration.
    To do that, we slightly delay the removal of lenses. More precisely, we define
    \[
    \lensfrozen_r(A):= \begin{cases}
            \lens_r(A) & r< \crit[A] \\[1.5ex]
            \lens_{\crit[A]}(A) & \crit[A]\leq r\leq (1+\epsilon)\crit[A] \\[1.5ex]
            \emptyset & (1+\epsilon)\crit[A] < r.
          \end{cases}
    \]
    We will sometimes call $\lensfrozen_r(A)$ a \emph{freezing lens} and $\lens_r(A)$ a \emph{non-freezing lens}.
    One can visualize the evolution of a freezing lens as a continuous process for increasing $r$: the lens $\lensfrozen_r$ grows until it reaches its freezing radius and remains unchanged (it is ``frozen'') for the interval $[\crit[A],(1+\epsilon)\crit[A]]$. Afterwards it completely disappears. We call $(1+\epsilon)\crit[A]$ the \emph{removal radius} of $A$. The construction is an adaptation of a similar one by Sheehy~\cite{Sheehy13-approxim-vietoris}.
    
    We write $\kcovfrozen_r$ for the union of $\lensfrozen_r(A)$ over all $A\in\binom{P}{k}$ and $\kcovfrozenfiltration:=\{\lensfrozen_r\}_{r\geq 0}$.
    
    \begin{lemma}\label{lem_frozen-lens-filtration} 
        $\kcovfrozenfiltration$ is a filtration, i.e., for any $r\leq r',$ $\kcovfrozen_r\subseteq\kcovfrozen_{r'}$.
    \end{lemma}
        
    \begin{proof}
        If the interval $(r,r']$ does not contain any removal radius,
        $\kcovfrozen_r\subseteq\kcovfrozen_{r'}$ because the inclusions hold lens-wise. Since the number of different removal radii is bounded by the number of sites and hence finite, it suffices to show that at a removal radius $s$, any lens that is removed is already covered by lenses that are not being removed at $s$.
        In fact, we show that such a lens is covered by lenses that are not yet frozen at $s$.
        
        Let $A$ be the $k$-subset associated with a lens being removed at $s$ and $x\in \lensfrozen_s(A)$.
        By definition, $s=(1+\epsilon)t$ with $t=\crit[A]$. 
        This implies that $x\in\lensfrozen_t(A)=\lens_t(A)\subseteq \kcov_t$ because the lens is frozen from radius $t$ on. 
        By Lemma~\ref{sandwich_first}, it follows that $x\in U_{(1+\epsilon)t}=U_{s}$, and therefore $x$ is contained in a lens $\lens_{s}(B)$ with $\crit[B]\geq s$.
        Thus $\lensfrozen_s(A)$ is covered by lenses $\lens_{s}(B)$ with $\crit[B]\geq s$ and $\kcovfrozenfiltration$ is a filtration.
    \end{proof}
    
    \begin{lemma} \label{lem:frozen_interleaving}
       $\kcovfrozenfiltration$ and $\kcovfiltration$ are $(1+\epsilon)$-interleaved.
    \end{lemma}
    
    \begin{proof}
        We show that $\kcovfrozen_r\subseteq\kcov_r\subseteq\kcovfrozen_{(1+\epsilon)r}$.
        The first inclusion is clear by definition. For the second inclusion, observe that $U_r\subseteq \kcovfrozen_r$
        follows directly from their definition. Then, Lemma~\ref{sandwich_first}
        yields $\kcov_r\subseteq U_{(1+\epsilon)r}\subseteq \kcovfrozen_{(1+\epsilon)r}$.
    \end{proof}
    
\section{A sparse simplicial filtration}\label{sec:simplicial_sparsification}
    Since $\lensfrozen_r(A)$ is closed and convex for every $r$, 
    the nerve of all (non-empty) $\lensfrozen_r(A)$ yields
    a simplicial complex homotopically equivalent the $k$-fold cover at radius $r$ because of the Nerve Theorem.
    However, the collection of simplicial complexes obtained when varying $r$ does not form a filtration
    because simplices disappear from the nerve when passing
    a removal radius.

 \subsection*{Cones.} 
\begin{figure}[ht!]
        \centering
        \includegraphics[scale=0.35]{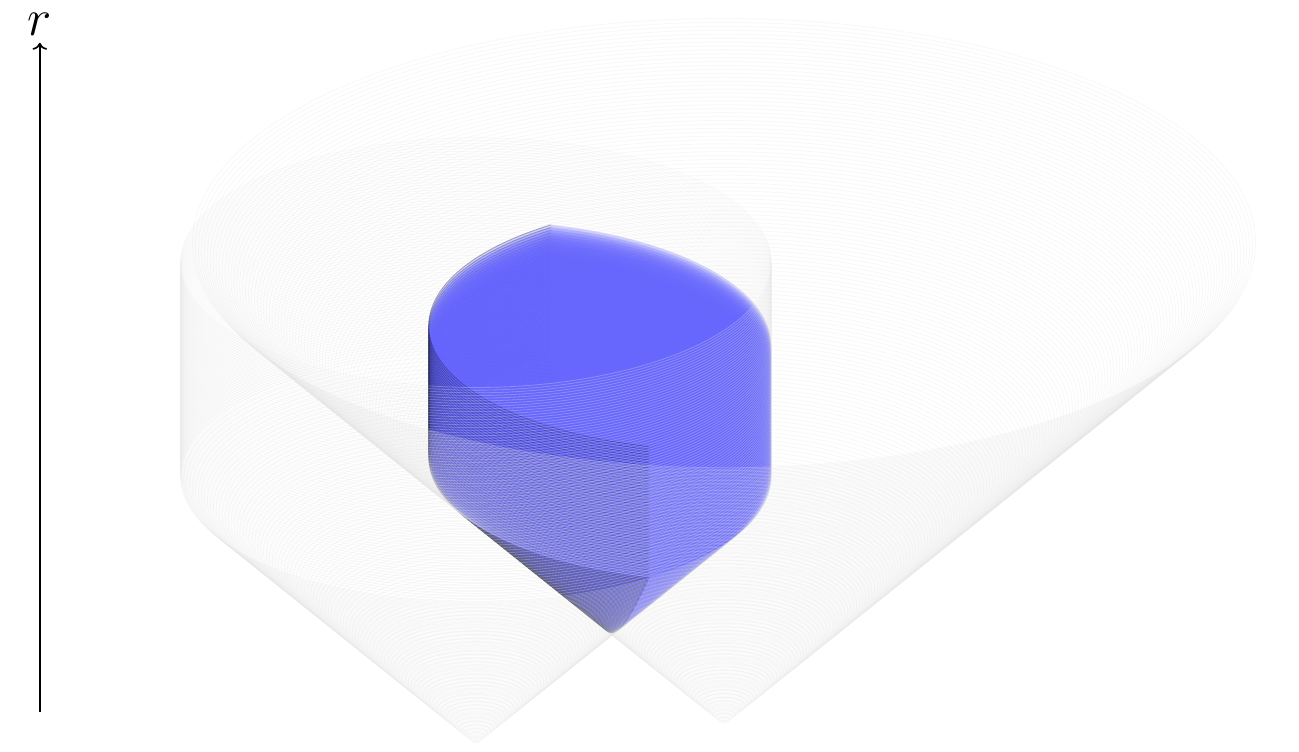}
        \caption{A cone representing the evolution of a lens for $k=2.$ At first the lens grows, until it is frozen. Then, the lens has static size and afterwards it disappears. At the radius where the lens disappears, it is completely covered by other lenses (which are not displayed in the figure).}
        \label{fig_cone-shape}
    \end{figure}

    We overcome the problem of removals of lenses by adapting a construction of Cavanna et al.~\cite{cavanna-2015-geometric_sparse}.
    The idea is to ``stack-up'' the lenses $\lensfrozen_r(A)$ for all radii:
    the \emph{cone} of $A$ at radius $r$ is
    \[
    \cone_r(A):=\underset{\alpha\in [0,r]}{\bigcup}\left(\lensfrozen_{\alpha}(A)\times \{\alpha\}\right)\subseteq \mathbb{R}^d\times\mathbb{R}.
    \]
    We write $\coneunion_r$ for the union of $\cone_r$ over all $A\in\binom{P}{k}$. Figure~\ref{fig_cone-shape} shows an example.
    
    \begin{lemma}\label{lem:stack_lemma}
    The filtrations $\conefiltration=\{\coneunion_r\}_{r\geq 0}$ and $\kcovfrozenfiltration$ are equivalent.
    \end{lemma}
    \begin{proof}
        In the same sense as above $\coneunion_r$ is a stacked-up version of $\kcovfrozen_\alpha$ for all $\alpha\leq r$, and we can consider $\kcovfrozen_r$
        as a subspace of $\coneunion_r$ via the map $x\mapsto (x,r)$ for $x\in\kcovfrozen_r$. Since $\kcovfrozenfiltration$ is a filtration, 
        there is a strong deformation retraction~\cite[Chap. 0]{hatcher_algebraic-topo} from $\coneunion_r$ to $\kcovfrozen_r$, given by $(x,\alpha),t\mapsto (x,(1-t)\alpha+tr)$,   
        which naturally commutes with the canonical inclusions. The result follows.
    \end{proof}
    
    We define the nerve of the cones as the \emph{sparse $k$th order \v{C}ech complex},
    \[
    \sparse_r:=\operatorname{Nrv}\left\{\cone_r\mid A\in\binom{P}{k}\right\}.
    \]
    $\sparse_r$ is a subcomplex
    of the $k$th order \v{C}ech complex for every $r$ because $\kcovfrozen_r\subseteq \kcov_r$.
    Moreover, if $r\leq r'$, $\cone_r(A)\subseteq\cone_{r'}(A)$ for all $A$. Hence $\sparsefiltration=\{\sparse_r\}_{r\geq 0}$ is a filtration. By the Persistent Nerve Theorem~\cite[Thm.~ 3.9]{alex_michael-2020-closed_nerve_thm} and Lemmas~\ref{lem:frozen_interleaving} and~\ref{lem:stack_lemma} we obtain:
    
    \begin{lemma}
        The filtrations $\sparsefiltration$ and $\conefiltration$ are equivalent.
        As a consequence, $\sparsefiltration$ is $(1+\epsilon)$-homotopy-interleaved with the $k$-fold filtration.
    \end{lemma}
    
 \subsection*{Discretization of the radius.}
    The filtration $\sparsefiltration$ is challenging to compute, due to the freezing of lenses.
    We elaborate on these issues in Section~\ref{sec:algorithm}. We now define a variant of $\sparsefiltration$ 
    which is easier to compute and still is an $(1+\epsilon)$-homotopy-interleaved with the $k$-fold filtration.
    
    Recall that the filtrations $\kcovfrozenfiltration$, $\conefiltration$ and $\sparsefiltration$
    are defined based on the freezing radii of sites, which depend on a parameter $\epsilon>0$.
    To obtain a $(1+\epsilon)$-approximation for $\epsilon\in(0,1]$ in the end, we consider the above construction of $\sparsefiltration$
    with parameter $\epsilon'=\frac{\epsilon}{3}$, obtaining a $(1+\frac{\epsilon}{3})$-homotopy-interleaving
    with the $k$-fold filtration.
    
    Next, for every $r\geq 0$, let $z\in\mathbb{Z}$ be such that $(1+\frac{\epsilon}{3})^z\leq r< (1+\frac{\epsilon}{3})^{z+1}$
    and define
    \[
    \discretesparse_r := \sparse_{(1+\epsilon/3)^z}.
    \]
    We call $\discretesparsefiltration:=\{\discretesparse_r\}_{r\geq 0}$ the \emph{discrete sparse $k$th order \v{C}ech filtration}. 
    It is formed by a discrete set of snapshots of $\sparsefiltration$ and kept unchanged except when passing over a snapshot radius
    (this is also referred to as the \emph{Left Kan extension} of a discrete filtration~\cite[Chap.~10]{MacLane_98_category-theory}).
    
    \begin{theorem}
        $\discretesparsefiltration$ is a $(1+\epsilon)$-homotopy-interleaving of the $k$-fold filtration.
    \end{theorem}
    \begin{proof}
        From the definition, $\discretesparse_r\subseteq\sparse_r\subseteq\discretesparse_{(1+\epsilon/3)r}$,
        so $\discretesparsefiltration$ is a $(1+\frac{\epsilon}{3})$-(homotopy)-interleaving of $\sparsefiltration$.
        Since $\sparsefiltration$ is a $(1+\frac{\epsilon}{3})$-homotopy-interleaving of the $k$-fold filtration, 
        we get that $\discretesparsefiltration$ is a $(1+\frac{\epsilon}{3})^2$-homotopy-interleaving of the $k$-fold filtration.
        The result follows by noting that $(1+\frac{\epsilon}{3})^2\leq 1+\epsilon$ for all $\epsilon\in (0,1]$.
    \end{proof}

\section{Size analysis}\label{sec:size_analysis}
    We bound the size of $\discretesparsefiltration$, i.e., the number of simplices it contains. Since $\discretesparse_r\subseteq\sparse_r$
    for all $r\geq 0,$ it is enough to bound the size of $\sparsefiltration$ with parameter $\epsilon'=\epsilon/3.$ Let 
    \[\infcone(A):=\bigcup_{r\geq 0}\cone_r(A)\]
    be the \emph{cone of $A$} (without dependence on a radius).
    Then, the size of $\sparsefiltration$ equals the size of the nerve of the cones $\infcone(A)$, where $A$ ranges over all $k$-subsets of sites.
    However, the number of vertices is not necessarily $\binom{n}{k}$ because many cones are empty:
    this happens in particular when the smallest radius for which the balls around the sites of $A$ intersect
    is larger than the removal radius of the lens. In fact, our argument shows that this is the case for the vast majority
    of cones. The proof for vertices extends readily to the case of $m$-wise intersections of cones, i.e. for $(m-1)$-simplices, without change and thus
    we treat the general case directly.
    
    For fixed $m\geq 1,$ we derive an upper bound for the number of sets $\{A_1,\ldots,A_m\}$
    such that the cones $\infcone(A_1),\ldots,\infcone(A_m)$ intersect. 
    Such sets are in one-to-one correspondence to the $(m-1)$-simplices of $\sparsefiltration$, hence we refer to these sets as $(m-1)$-simplices. 
    Let $\sigma=\{\infcone(A_1), \ldots, \infcone(A_m)\}$ be such a $(m-1)$-simplex and the set of sites $P$ be ordered according to the \permut.
    We call a site $p_i$ \emph{involved} in $\sigma$ if $p_i$ belongs to one of the sets $A_1,\ldots,A_m$.
    Note that there are at most $km$ sites involved in $\sigma$.  We say that $\sigma$ is \emph{associated} to a site $p_i$ if $p_i$ is involved in $\sigma$ and all other involved sites have index smaller than $i$. 
    Our strategy is to upper bound the number of $(m-1)$-simplices in $\sparsefiltration$ associated to an arbitrary $p_i$. 
    
    Fix $i\geq k+1$ and $\omega_i:=(1+\epsilon')\crit[p_i]<\infty$. 
    Let $B$ denote the ball of radius $2\omega_i$ centered at $p_i$.
    
    \begin{lemma}\label{lem:simple_ball_lemma}
    If $\sigma:=\{\infcone(A_1),\ldots,\infcone(A_m)\}$ is an $(m-1)$-simplex in $\sparsefiltration$ associated to $p_i$, then all sites involved in $\sigma$ are contained in $B$.
    \end{lemma}
    \begin{proof}
        Let $\alpha$ denote the minimal radius such that all the balls around sites involved in $\sigma$ intersect. This is the radius of the minimum enclosing ball of the involved sites.
        Any common intersection of the cones must happen at scale $r\geq\alpha.$
        
        On the other hand, assume wlog that $p_i\in A_1$. Since $p_i$ has maximal index in $A_1$, we have that $\crit[A_1]=\crit[p_i]$. Hence the removal radius of $A_1$ is equal to $\omega_i$ and it follows that the cone of $A_1$ is empty for all radii greater than $\omega_i$.
        Therefore any common intersection of the cones of $\sigma$ must happen at scale $r\leq \omega_i$.
        Hence, as we assume that the cones do intersect, we must have that $\alpha\leq\omega_i$.
        It follows that for any involved site $q$, $\operatorname{d}(q,p_i)\leq 2\alpha\leq 2w_i$.
        \end{proof}
    
    Hence the involved sites of $\sigma$ are close to $p_i$ in the sense of the lemma. We can furthermore guarantee that the points of $P_i$ are not too densely packed in $B$ using the packing property of the \permut.
    
    \begin{lemma}\label{lem:packing_sites}
        The ball $B$ contains at most
        $\hypertarget{link_sites-bound}{\numberofsites} := k\left(\dfrac{96}{\epsilon}\right)^{\delta}$
        sites of $P_i=\{p_1,\ldots,p_i\}$, where $\delta$ is the doubling dimension of $P$.
    \end{lemma}
    \begin{proof}    
        We cover $B\cap P$ by balls of radius ${\lambda_{i}}/{4}$. That can be done with at most
        $\zeta=\left(\dfrac{16(1+\epsilon')^2}{\epsilon'}\right)^\delta$
        balls (see \emph{Doubling Dimension} in Section~\ref{sec:background}). By the Packing Lemma~\ref{lem_packing}, each open ball of radius $\lambda_{i}/4$ contains at most $k$ sites of $P_i$, thus the total number of sites in $B$ is at most $k\zeta$. The bound follows because $\epsilon'=\frac{\epsilon}{3}$ and $\epsilon\leq 1$, hence
        $(1+\epsilon')^2\leq\frac{16}{9}<2$.
    \end{proof}
    
    Bounding the number of non-empty $m$-intersections of cones is now a matter of simple combinatorics. 
    \begin{theorem}
\label{thm:size_bound}
    The number of $(m-1)$-simplices of $\sparsefiltration$ is at most
    \[
    n\cdot\numberofsites^{km}=nk^{km}\left(\dfrac{96}{\epsilon}\right)^{\delta k m}.
    \]
    \end{theorem}
    
    \begin{proof}
    Fix $p_i$ with $i\geq k+1$. Every $(m-1)$-simplex associated to $p_i$ with non-empty cone intersection has up to $km$ involved sites,
    which all lie in $B$ by Lemma~\ref{lem:simple_ball_lemma}. Moreover, all involved sites are in $P_i$ and, by Lemma~\ref{lem:packing_sites},
    there are at most $\numberofsites$ of those sites in $B$ to choose from. It follows that there are at most $\numberofsites^{km}$ different choices
    possible. This upper bound holds for every $p_i$ with $i\geq k+1$, so multiplying by the number of sites $(n-k)$ yields a bound of
    $(n-k)\numberofsites^{km}$ for simplices associated to $p_i$ with $i\geq k+1$.

    Finally, note that no simplex gets associated to $p_i$ for $i=1,\ldots,k-1$, and the only simplex associated to $p_k$ is the vertex defined by $\{p_1,\ldots,p_k\}$.
    Hence the total size bound is $(n-k)\numberofsites^{km}+1$, and since $\numberofsites\geq 1$ and $k\geq 1$, the result follows.
    \end{proof}
    
    We remark that the bounds on this section are not tight and slightly better ones could be easily achieved, by keeping binomials in place or avoiding some approximations. However the improvements would be minor.
    
\section{Computation}\label{sec:algorithm}
    We now present an algorithm to construct the discrete sparse \v{C}ech filtration. As in the previous sections, let us fix a finite set $P\subseteq\mathbb{R}^d$, an integer $k>0$
    and $\epsilon\in(0,1]$. Assume that $P=\{p_1,\ldots,p_n\}$ has the indices ordered with respect to the \permut, and that we have computed the corresponding values $\lambda_1,\ldots,\lambda_n$ as discussed in Section~\ref{sec:k-dist_permut}.
    The algorithm outputs the discrete sparse $k$th order \v{C}ech filtration $\discretesparsefiltration=\{\discretesparse_r\}_{r\geq 0}$ as a list of simplices with their corresponding \emph{critical value}, i.e., the smallest parameter value $r$ 
    for which the simplex is part of the filtration. Note that by definition of the discrete sparse \v{C}ech filtration, every critical value is of the form $(1+\epsilon/3)^z$ for some integer $z$.
Our algorithm follows the approach and notation of Section~\ref{sec:size_analysis}. 
    
 \subsection*{Friends and vertices}
    The first step is to find, 
    among $p_1,\ldots,p_{i-1}$, all sites of distance at most $2\omega_i$ from $p_i$,
    where $\omega_i=(1+\epsilon')\crit$ (compare Lemma~\ref{lem:simple_ball_lemma}).
    We call these points \emph{friends} of $p_i$. 
    
    We compute friends using the quadtreap data structure introduced in Section~\ref{sec:background}: 
    we initialize the quadtreap with the first $k$ sites $p_1,\ldots,p_k$ of the permutation.
    For every site $p_{k+1},\ldots,p_n$ in order, we query the quadtreap for $p_i$ with radius $2\omega_i$,
    obtaining a candidate list $F_i$ of friends of $p_i$. We traverse the list $F_i$ and remove
    those ``false friends'' $x$ from the list for which $d(p_i,x)>2\omega_i$.
    Then, we add $p_i$ to the quadtreap.

    It is straight-forward to see that the (expected) cost of one iteration is $O(\log n+|F_i|)$.
    All points in $F_i$ have distance at most $4\omega_i$ from $p_i$ by the property of the quadtreap,
    and by the same argument as in the proof of Lemma~\ref{lem:packing_sites}, 
    the number of sites reported for $p_i$ is at most $k\left(\dfrac{192}{\epsilon}\right)^\delta=\mathcal{O}(k(1/\epsilon)^\delta)$.
    Thus  we get the friends of $p_i$
    for all sites $p_i$ in expected time 
    $\mathcal{O}(n\log n + nk(1/\epsilon)^\delta)$.
    Note that the number of friends is bounded by $\hyperlink{link_sites-bound}{\numberofsites}$ as defined in Lemma~\ref{lem:packing_sites}.
   
    The next step is to compute the vertices associated to $p_i$ that enter the filtration, for each $i\geq k$. 
    We proceed by brute-force, just enumerating all $k$-tuples formed by $p_i$ and $k-1$ of its friends and checking for every $k$-tuple whether their cone is non-empty. The last condition is simple to check, as the cone is non-empty if and only if the radius $\alpha$ of the minimum enclosing ball of the $k$ sites
    is at most $\crit$. In this case, the critical radius of the vertex is set to $(1+\epsilon/3)^z$, where $z$ is the smallest integer such that $(1+\epsilon/3)^z\geq \alpha$.
    Computing $\alpha$ and $z$ per vertex requires $\mathcal{O}(k)$ in expectation~\cite[Chap. 4.7]{2008_dutch-book_algorithms}. 
    Hence we calculate all vertices in $\mathcal{O}((n-k) k \numberofsites^{k-1})$ time. We summarize:
\begin{lemma}\label{lem:friends_and_vertices}
We can compute all friends of sites and vertices of $\discretesparsefiltration$ in expected time
\[ \mathcal{O}(n\log n + nk(1/\epsilon)^\delta+(n-k) k \numberofsites^{k-1}).\]
\end{lemma}

 \subsection*{Intersections of cones}
     To treat the case of higher-dimensional simplices, we discuss the major predicate required:
     Given subsets $A_1,\ldots,A_{m+1}$ of sites, decide whether their cones $(\infcone(A_1),\ldots,\infcone(A_{m+1}))$
     intersect, and if they do, compute the smallest radius $r$ for which they intersect.
     Answering this question is technically challenging because the cones might first intersect for a radius where one or several of the associated lenses are frozen.
     Therefore, unlike for the question whether a single cone is empty or not, one cannot resolve this question in general by a simple minimal enclosing ball computation on the involved sites.
    
    Nevertheless, that minimal enclosing ball of the involved sites yields partial information.
    Let $\mebr$ denote the radius of the minimum enclosing ball of the sites in $A_1\cup\ldots \cup A_{m+1}$. 
    Let $\Lambda$ denote the smallest freezing time among $\crit[A_1],\ldots,\crit[A_{m+1}]$.
    There are two simple cases.
    
    \begin{description}
        \item [$\mebr\leq\Lambda$:] In this case, $\lensfrozen_{\mebr}(A_i)=\lens_{\mebr}(A_i)$ for all $i$ and they intersect at radius $\mebr$. 
        This means that $\sigma$ belongs to the filtration and its critical value is the minimal value of the form $(1+\epsilon')^z$ that is greater or equal to $\mebr$.
        \item [$(1+\epsilon')\Lambda\leq\mebr$:] Here, the non-freezing lenses only intersect for a radius greater than the removal radius of one lens. 
        Since the freezing lenses are subsets of their non-freezing counterparts, it follows that $\lensfrozen_r(A_1),\ldots,\lensfrozen_r(A_{m+1})$ do not intersect before one of them becomes empty. Hence $\sigma$ does not belong to the filtration in this case.
    \end{description}
    
    The remaining and most challenging case is if $\Lambda<\mebr\leq (1+\epsilon')\Lambda$. To manage this case, we first consider a more general geometric problem: 
    given balls $\beta_1,\ldots,\beta_\ell$ with centers $x_1,\ldots,x_\ell$ and radii $r_1,\ldots,r_\ell$, decide whether $\beta_1\cap\ldots\cap \beta_\ell\neq\emptyset$ \hypertarget{star}{$(\star)$}.
    Assume without loss of generality that all radii $r_1,\ldots,r_\ell$ lie between $0$ and $1$
    (otherwise, scale the instance). Now, define $\beta_{j}'$ as the ball with center $x_j$ and radius $1-r_j$, for $1\leq j\leq \ell$. We use the following geometric predicate to deal with this problem.
        
    \begin{lemma}
    \label{lem:common_intersection}
            The balls $\beta_1,\ldots,\beta_\ell$ have a common intersection point if and only if there exists a ball of radius $1$ that contains $\beta_1',\ldots,\beta_\ell'$.
    \end{lemma}
    \begin{proof}
            Assume there is a ball $B_1(c)$ that contains each $\beta_{j}'$ . Then $d(c,x_j)+(1-r_j)\leq 1$ and thus $d(c,x_j)\leq r_j$ for each $1\leq j\leq\ell$. It follows that $c\in \beta_{j}$ for every $1\leq j\leq\ell$, implying ``$\Leftarrow$''. The argument can be reversed to show the equivalence.
    \end{proof}
    
    Next, consider the following question: given $A_1,\ldots,A_{m+1}$ and a fixed radius $r$,
    decide whether the corresponding cones intersect at level $r$, or equivalently, whether
    $\lensfrozen_r(A_1)\cap\ldots\cap\lensfrozen_r(A_{m+1})\neq\emptyset$.
    We can easily reduce this problem to \hyperlink{star}{$(\star)$}, since the intersection is simply an intersection
    of (up to $k(m+1)$) balls whose radius is either $r$ or $\crit[A_i]$, and we can compute this radius
    easily since we know $\crit[A_i]$. We refer to this predicate as \emph{checking the cones of $A_1,\ldots,A_m$
    at radius $r$}.
    
    Now, we handle the case of $\Lambda<\mebr\leq (1+\epsilon')\Lambda$ as follows: We first check the cones
    at radius $(1+\epsilon')\Lambda$.
    If there is no intersection, $\sigma$ does not belong to the filtration.
    If there is an intersection, $\sigma$ is added to the filtration.
    To determine its critical value, let $z$ be the smallest integer such that $(1+\epsilon')^{z}\geq\Lambda$.
    Check the cones at radius $(1+\epsilon')^z$. If they intersect, set the critical value of $\sigma$ to be $(1+\epsilon')^{z}$. 
    Otherwise, set it to $(1+\epsilon')^{z+1}$. This ends the description of the algorithm to determine whether $\sigma$
    belongs to the filtration.
    
    We argue about the correctness: first, note that $(1+\epsilon')\Lambda$ is the maximum value for which no freezing lens is empty.
    This implies that if the cones are intersecting at all, they intersect at level $(1+\epsilon')\Lambda$.
    Hence, checking the cones at $(1+\epsilon')\Lambda$ decides whether $\sigma\in\discretesparsefiltration$ or not.
    
    If $\sigma\in\discretesparsefiltration$, by definition of the filtration,
    the critical value is necessarily an integer power of $(1+\epsilon')$. 
    Note that $(1+\epsilon')^{z+1}\geq (1+\epsilon')\Lambda$, so the cones
    $(C_{(1+\epsilon')^{z+1}}(A_1),\ldots,C_{(1+\epsilon')^{z+1}}(A_m))$
    intersect, implying that the critical value is at most $(1+\epsilon')^{z+1}$.
    On the other hand, by the choice of $z$, $(1+\epsilon')^{z-1}<\Lambda<\mebr$, where the last inequality
    comes from our case distinction. It follows that the non-freezing lenses at radius $(1+\epsilon')^{z-1}$
    do not intersect, and neither do the freezing lenses because they are subsets of their non-freezing counterparts. 
    This implies that the critical value of $\sigma$
    is larger than $(1+\epsilon')^{z-1}$. Combing these two results, we obtain that the critical value
    is either $(1+\epsilon')^{z}$ or $(1+\epsilon')^{z+1}$, and the algorithm decides this by checking the cones
    at $(1+\epsilon')^{z}$.
    
    The above procedure requires to compute one minimum enclosing ball (to compute $\mebr$), and two checks of cones at different radii,
    which reduce to two instances of problem $(\star)$.
    Problem $(\star)$, in turn, can be answered by computing the \emph{minimum enclosing ball of balls}~\cite{Fischer_Gartner-2004-MEB_of_balls}
    using Lemma~\ref{lem:common_intersection}. 
    For constant dimension, the expected running time for computing the minimum enclosing ball of $\ell$ balls is $\mathcal{O}\left(\ell\right)$, 
    the same as for standard minimum enclosing ball computations.
    Hence, we obtain
    
    \begin{lemma}\label{lem:cone_intersection_predicate}
        Given $\sigma=(A_1,\ldots,A_m)$, deciding whether $\sigma\in\discretesparsefiltration$ and determining its critical value
        can be done in $\mathcal{O}\left(km\right)$.
    \end{lemma}
    
    We point out that we were not able to find an algorithm to compute the critical value in $\sparsefiltration$ instead of $\discretesparsefiltration$, 
    which is the reason why we introduced the latter in Section~\ref{sec:simplicial_sparsification}.
    One could approximate the smallest $r$ in $\sparsefiltration$ arbitrarily closely by combining the minimum enclosing ball method described above and a binary search, 
    but this appear to make the approximation scheme more complicated, and the result would still only approximate $\sparsefiltration$,
    as our approach does.

 \subsection*{Higher dimensional simplices} 
    We now describe how to determine the higher-dimensional simplices associated to $p_i$ in an output-sensitive fashion.
    We proceed inductively by dimension, up to a maximal dimension $\maxdim$.
    Fix a $(m-1)$-simplex $\sigma'=\{A_1,\ldots,A_m\}$  associated to $p_i$. We compute all cofacets of $\sigma'$ in the filtration, that is, all 
    $m$-simplices that contain $\sigma'$ and one further vertex $A_{m+1}$. 
    Note that we could avoid some repetitions by fixing a global order on the $k$-subsets and only check tuples $(A_1,\ldots,A_{m+1})$ that are
    increasing in this order, but the saving would be negligible in terms of complexity.
    
    Since each element of $A_{m+1}$ is either a friend of $p_i$ or $p_i$ itself,
    we enumerate all $k$-subsets consisting of these sites.
    Writing $A_{m+1}$ for such a subset, we check whether the cones of $(A_1,\ldots,A_m,A_{m+1})$
    intersect, using the algorithm of the previous subsection. If yes, we add the corresponding
    $m$-simplex with its critical value to $\discretesparsefiltration$. 
    Note that if that happens, $A_{m+1}$ is necessarily a vertex of $\discretesparsefiltration$,
    as otherwise, the cone of $A_{m+1}$ is empty
    and so is the intersection, of the cones of $(A_1,\ldots,A_{m+1})$.
    
    The number of $k$-subsets consisting of $p_i$'s friends is bounded by $\binom{\numberofsites}{k}$, where we recall that $\hyperlink{link_sites-bound}{\numberofsites}$ bounds the number of friends. 
    Therefore, using Lemma~\ref{lem:cone_intersection_predicate}, the expected running time spent checking for cofacets is 
    $\mathcal{O}(mk\numberofsites^k)$ per $(m-1)$-simplex of the filtration. 
    Doing this over all simplices of the filtration, up to dimension $\maxdim$, yields a total expected complexity of 
        \[\mathcal{O}(X\cdot(\maxdim+1) \cdot k\cdot \numberofsites^k)\]
    for the step of computing higher dimensional simplices, where $X$ is the total number of simplices in the filtration. 
    Recalling the bound of Lemma~\ref{lem:friends_and_vertices} for the complexity for computing friends and vertices, we obtain a total complexity of
        \begin{equation}\label{chap1_eq:running_time_construc}
            \mathcal{O}\left(n\log n + \frac{nk}{\epsilon^\delta}+(n-k) k \numberofsites^{k-1}+X\cdot (\maxdim+1)\cdot k\cdot \numberofsites^k\right)
        \end{equation}
    for computing all simplices of $\discretesparsefiltration$ and their critical values, assuming that the \permut is computed.
    
    We next show that the second and third summand in the bound are dominated by the last term. We first show a simple lower bound for $X$:
    
    \begin{lemma}
    \label{chap1_lem:output_simplices_lower_bound}
            The number of output simplices is $X\geq n-k+1$.
    \end{lemma}
    \begin{proof}
    For every $p_i$ with $i\geq k$, consider the $k$-subset $A_i$ consisting of $p_i$ and its $(k-1)$-closest neighbors in $P_{i-1}$.
    Then $A_k$ is the $k$-subset $\{p_1,\ldots,p_k\}$ which is the unique vertex of $\discretesparsefiltration$ that does not freeze.
    For $i>k$, the distance of $p_i$ to every site in $A_i$ is at most $d^{k-1}(p_i,P_{i-1})\leq d^k(p_i,P_{i-1})=\lambda_i$.
    That means that $\lens_{\lambda_i}(A_i)$ is not empty.
    On the other hand, $\crit[A_i]=\crit[p_i]=\frac{(1+\epsilon')\lambda_i}{\epsilon'}>\lambda_i$, so the lens is not frozen until $\lambda_i$.
    This implies that $A_i$ is a vertex of $\discretesparsefiltration$, and since $A_{k},\ldots,A_n$ are pairwise disjoint, the bound follows.
    \end{proof}
    
    The lemma implies immediately that the third term is dominated by the fourth. For the second term, note that $n\leq 2(n-k+1)k$
    by an elementary argument, hence $\frac{nk}{\epsilon^\delta}\leq (n-k+1)k^2\frac{2}{\epsilon^\delta}\leq Xk\numberofsites$,
    since $\numberofsites\geq k\frac{2}{\epsilon^\delta}$.

    Hence, (\ref{chap1_eq:running_time_construc}) together with the algorithm from Section~\ref{sec:k-dist_permut} yields the final result: 

    \begin{theorem}
    \label{thm:computation_time}
    Given a set $P$ of $n$ points, $k\geq 1$ and $\epsilon\in(0,1]$, the discrete sparse $k$th order \v{C}ech filtration $\discretesparsefiltration$
    up to dimension $\maxdim$ can be computed in time
    \[\mathcal{O}\left(n k \log n\log\spread+X k^{k+1} (\maxdim+1) \left(\dfrac{96}{\epsilon}\right)^{k\delta}\right),\]
    where $\spread$ is the spread of $P$ and $X$ is the total number of simplices in the filtration.
    \end{theorem}

\section{Approximating multicovers}
\label{sec:multicovers}
    We remove the restriction of $k$ being a fixed parameter.
    Let us write  $\kcov^{(k)}_r$ instead of $\kcov_r$ for the $k$-fold cover,
    to emphasize the dependence on $k$. From the definition of lenses, it holds that
    $A\subseteq B$ implies $\lens_r
    (B)\subseteq\lens_r
    (A)$ and hence
    $\kcov^{(k')}_{r'}\subseteq\kcov^{(k)}_{r}$
    whenever $k'\geq k$ and $r'\leq r$. 
    The collection of $\kcov^{(k)}_r$ for $1\leq k\leq n$ and $r\geq 0$ is called
    the \emph{multicover} of a point set. For a fixed $\maxk\geq 1$, if we restrict
    the values of $k$ between $1$ and $\maxk$, we call the collection
    the \emph{multicover up to $\maxk$}.

    To define and compute a simplicial approximation of the multicover up to $\maxk$, we need to extend some further definitions of Section~\ref{sec:background}.
    A collection of topological spaces
    $\mathcal{X}=(X_{r,k})_{r\geq 0,k\in\{1,\ldots,\maxk\}}$
    is called a \emph{bifiltration} (with index set $\mathbb{R}^+_0\times\{1,\ldots,\maxk\}$)
    if whenever $k'\geq k$ and $r'\leq r$, $X_{r',k'}\subseteq X_{r,k}$.
    The multicover up to $\maxk$ is a bifiltration.
    We call two bifiltrations $\mathcal{X}$ and $\mathcal{Y}$ \emph{(homotopy) equivalent}
    if there is a family of maps $\{f_{r,k}:X_{r,k}\to Y_{r,k}\}_{r\geq 0,k\in\{1,\ldots,\maxk\}}$
    that are homotopy equivalences of spaces and additionally commute with the inclusion maps
    of $\mathcal{X}$ and $\mathcal{Y}$. 
    We call two bifiltrations  $\mathcal{X}$ and $\mathcal{Y}$ 
    \emph{$(1+\epsilon)$-interleaved} if for every $(r,k)$, we have $Y_{r,k}\subseteq X_{r,k}\subseteq Y_{(1+\epsilon)r,k}$;
    in other words, the restricted filtrations for $k$ fixed are $(1+\epsilon)$-interleaved.
    We call $\mathcal{X}$ and $\mathcal{Y}$ \emph{$(1+\epsilon)$-homotopy\--interleaved} if there exist bifiltrations $\mathcal{X}'$
    equivalent to $\mathcal{X}$ and $\mathcal{Y}'$ equivalent to $\mathcal{Y}$ such that $\mathcal{X}'$ and $\mathcal{Y}'$
    are $(1+\epsilon)$-interleaved. We remark that this notion of $(1+\epsilon)$-(homotopy-)interleaving is stronger than
    the usual definition of interleaving~\cite{Chazal-2009-interleavings}. Moreover, it implies
    that the induced $2$-parameter persistence modules of $\mathcal{X}$ and $\mathcal{Y}$ are $(1+\epsilon)$-interleaved,
    which is the common notion of closeness of two data sets in the context of multi-parameter persistence.
    
    Given a point set $P$, this section aims
    to define a simplicial bifiltration of small size that is an $(1+\epsilon)$-homotopy-interleaved
    with the multicover up to $\maxk$ of $P$.
    
 \subsection*{Freezing lenses revisited.}
    Recall that our approach in Section~\ref{sec:lenses_sparsification} (for fixed $k$) yields the 
    filtration $\kcovfrozenfiltration^{(k)}$ that is $(1+\epsilon)$-interleaved with $\kcovfiltration^{(k)}$.
    However, $\left\{\kcovfrozen_r^{(k)}\right\}_{r\geq 0,k\in\{0,\ldots,\maxk\}}$ does \emph{not} form a bifiltration, because
    for $k'\geq k$ and $r\geq 0$, $\kcovfrozen^{(k')}_r\subseteq\kcovfrozen^{(k)}_r$ does not hold in general.
    The reason is that the freezing values of $\kcovfrozen^{(k)}$ are based on the $k$-distance permutation
    and those of $\kcovfrozen^{(k')}$ on the $k'$-distance permutation. There is no evident relationship
    between these two permutations (even if the initial points are picked in a compatible way):
    sites that appear early in the permutation for $k$ can appear late in the permutation for $k'$ and vice versa,
    so that lenses can freeze at arbitrary radii. We remark that the inclusion ``almost'' holds,
    in the sense that $\kcovfrozen^{(k')}_r\subseteq\kcovfrozen^{(k)}_{(1+\epsilon)r}$ 
    for $k'\geq k$, but the technical implications of this approximation appear to be non trivial to handle.
    
    Fortunately, there is an easy fix for this issue: we compute the $\maxk$-distance permutation
    (where $\maxk$ is the upper bound of $k$ from the problem statement)  and define the freezing times
    of each $p_i$ with respect to the $\lambda_i$ obtained from this permutation.
    In particular, in the definition of $\lensfrozen_r(A)$, the value $\crit[A]$
    is independent of the cardinality of $A$ and for $A\subseteq B$ subsets of sites, we obtain that 
    $\lensfrozen_r(B)\subseteq \lensfrozen_r(A)$. 
    This yields immediately that the redefined $\kcovfrozenfiltration:=\left\{\kcovfrozen_r^{(k)}\right\}_{r\geq 0,k\in\{0,\ldots,\maxk\}}$ is a bifiltration.
    
    Moreover, the redefined $\kcovfrozenfiltration$ is still $(1+\epsilon)$-interleaved with the multicover up to $\maxk$.
    To see that, note that the only property of the distance permutation required to establish the sandwiching
    in Section~\ref{sec:lenses_sparsification} was the covering property (Lemma~\ref{lem_covering}), used
    in Lemma~\ref{sandwich_first}. The covering property, however, is unaffected when using the $\maxk$-distance
    permutation instead of the $k$-distance permutations for $k\leq\maxk$. The proof of the following lemma
    is verbatim the same as the proof of Lemma~\ref{lem_covering}, adding the fact that $d^k(p,A)\leq d^\maxk(p,A)$
    for $k\leq\maxk$, all sites $p\in P$ and all subsets $A$.
    \begin{lemma}[Generalized Covering]
        Let $P_1,\ldots,P_n$ and $\lambda_1,\ldots,\lambda_n$ be determined by the $\maxk$-distance permutation,
        and $k\leq\maxk$. Let $\kcov^{(k)}_r|_{P_i}$ denote the $k$-fold cover of $P_i$. Then
        \[
        \kcov^{(k)}_r|_{P_i}\subseteq\kcov^{(k)}_r \subseteq\kcov^{(k)}_{{r+\lambda_{i+1}}}|_{P_i}.
        \]
    \end{lemma}
    With this generalization, all results in Section~\ref{sec:lenses_sparsification} carry over using the redefined
    freezing times of sites.
    
    \subsection*{Cones revisited.}
    Recall that we constructed our simplicial approximation by ``stacking up'' the freezing lenses $\lensfrozen_{\alpha}(A)$ with $\alpha\in[0,r]$ 
    to a cone $\cone_r(A)$
    in one dimension higher; and considering the nerve $\sparse_r^{(k)}$ of all cones over $k$-subsets of sites. 
    Even with the freezing lenses redefined
    as above, the resulting simplicial complexes do not form a bifiltration, since for $k\neq k'$, already the vertex sets
    of $\sparse_r^{(k)}$ and $\sparse_r^{(k')}$ are disjoint, being $k$-subsets and $k'$-subsets of sites, respectively.
    
    We suggest a very simple fix. For $k\leq\maxk$, we simply redefine
    \[
    \sparse_r^{(k)}:=\operatorname{Nrv}\left\{\cone_r(A)\,\,\middle\vert\,\, A\in\binom{P}{\ell},k\leq\ell\leq\maxk\right\}.
    \]
    In other words, we take the nerve of the cones over all subsets with at least $k$ sites (up to $\maxk$).
    On the one hand, it is now clear that $\sparse_r^{(k)}$ is a subcomplex of $\sparse_r^{(k')}$ for $k\leq k'$, just because
    the set of cones in the former nerve is a subset of the cones in the latter nerve.
    Hence, the complexes $\sparse_r^{(k)}$ form a bifiltration which we denote again by $\sparsefiltration$.
    On the other hand, the bifiltration $\sparsefiltration$ is equivalent to $\kcovfrozenfiltration$: the Persistent Nerve Theorem~\cite{alex_michael-2020-closed_nerve_thm}
    also applies to bifiltrations and asserts that the union of the cones in the definition
    of $\sparse_r^{(k)}$ is equivalent to the nerve. Moreover for $A\subseteq B$, the cone $\cone_r(B)$ is contained in $\cone_r(A)$
    (because the freezing lenses are included for every $r$), thus the union of cones in $\sparse_r^{(k)}$ does not change
    from the old to the new definition. 
    It follows that the redefined bifiltration $\sparsefiltration$ is equivalent to $\kcovfrozenfiltration$,
    and therefore is 
    $(1+\epsilon)$-homotopy-interleaved with the multicover up to $\maxk$. We can pass to the discretized version $\discretesparsefiltration$
    taking snapshots at every $(1+\frac{\epsilon}{3})^z$ for all $k=1,\ldots,\maxk$ and using the Left Kan extension, just as in Section~\ref{sec:simplicial_sparsification}.
    
 \subsection*{Size bounds and computation revisited.}
    Our redefined simplicial approximation is the nerve of a much larger cover, which mixes cones of different cardinalities.
    Perhaps surprisingly, the size analysis and the derived complexity bounds remain basically the same, replacing $k$ with $\maxk$:

    \begin{theorem}
        The number of $(m-1)$-simplices that appear in the simplicial bifiltration $\discretesparsefiltration$ is at most
        \[
        n\cdot\numberofsites^{\maxk m}=n\maxk^{\maxk m}\left(\dfrac{96}{\epsilon}\right)^{\delta \maxk m}.
        \]
    \end{theorem}
    \begin{proof}
        Considering an $(m-1)$-simplex $\sigma:=\{\infcone(A_1),\ldots,\infcone(A_m)\}$, with all $A_i$ of cardinality
        between $1$ and $\maxk$, we have up to $\maxk m$ involved sites. Associating such a simplex to the site $p_i$
        with largest index in the permutation, we have two cases: if $\crit[p_i]$ is infinite, the involved sites
        must be selected from the first $\maxk$ sites in the $\maxk$-distance permutation, and there are at most $\maxk^{\maxk m}$
        different choices for that.
        If $\crit[p_i]$ is finite, we proceed as in Section~\ref{sec:size_analysis}: every involved site must be in a ball
        of radius $2\crit[p_i]$ around $p_i$, and the packing property of the $\maxk$-distance permutation allows us 
        to bound the number of sites in that call with index smaller than $i$ by the same $\numberofsites$ as in Lemma~\ref{lem:packing_sites},
        just with $k$ replaced with $\maxk$. The combinatorial argument in the proof of Theorem~\ref{thm:size_bound} carries over
        and we obtain a total bound of
        \[\maxk^{\maxk m}+(n-\maxk)\cdot\numberofsites^{\maxk m}\]
        where the $(n-\maxk)$ in the second summand comes from the fact that $(n-\maxk)$ sites $p$ satisfy $\crit[p]<\infty$.
        Since $\numberofsites\geq\maxk$, the bound follows.
    \end{proof}
    
    The computation of $\discretesparsefiltration$ can be easily adapted to the redefined objects:
    the bifiltration $\discretesparsefiltration$ is \emph{$1$-critical}, which means that
    for every simplex $\sigma$ in the bifiltration, there is a unique critical value $(r_0,k_0)$
    such that $\sigma\in D_r^{k}$ if and only if $r\geq r_0$ and $k\leq k_0$.
    For a collection of cones $\infcone(A_1),\ldots,\infcone(A_m)$, with every $A_i$ being of
    cardinality between $1$ and $\maxk$, the main task is to check whether the common intersection
    is non-empty (in which case they form a $(m-1)$-simplex in $\discretesparsefiltration$), 
    and in this case to compute the critical value.
    The intersection predicate from Section~\ref{sec:algorithm} carries over because it does
    not depend on the number of (weighted) balls considered. The same is true for determining
    the smallest value of $r_0$ where they intersect. Writing $k_0$ for the smallest cardinality
    among the sets $A_1,\ldots,A_m$, it is then clear that the corresponding simplex appears
    in the bifiltration exactly for all $k\leq k_0$ (as for values 
    larger than $k_0$, one of the sets $A_i$ is missing),
    and $(r_0,k_0)$ is the critical value. Other adaptions of how to traverse the cones
    to obtain an output-sensitive bound are straight-forward and we omit the details. Finally, we arrive at
    
    \begin{theorem}
    \label{thm:computation_time_multi}
            Given a set $P$ of $n$ points, $m,\maxk\geq 1$ and $\epsilon\in(0,1]$, our algorithm computes all simplices
    of dimension up to $(m-1)$
            of a simplicial $(1+\epsilon)$-approximation of the multicover of $P$ up to $\maxk$ in time
            \[\mathcal{O}\left(n \maxk \log n\log\spread+X \maxk^{\maxk+1} m \left(\dfrac{96}{\epsilon}\right)^{\maxk\delta}\right),\]
            where $\spread$ is the spread of $P$ and $X$ is the total number of simplices in the filtration.
    \end{theorem}

\section{Conclusions}\label{sec:conclusion}
    We introduced the first $(1+\epsilon)$-approximate filtration of the higher order $k$-fold filtration and provided an algorithm for computing it. If $k$ and $\epsilon$ are considered as constants and the input point set has constant spread, the algorithm runs in time $\mathcal{O}(n\log n)$ and yields a filtration of size $\mathcal{O}(n)$, which are the same favorable properties as in the well-studied case $k=1$.

    There are various avenues to strengthen and generalize our results. First of all, our method has concentrated on the Euclidean case, but our approach mostly
    generalizes to point sets in arbitrary metric spaces. The algorithm cannot use the quadtreap data structure anymore in this case, but there is no need for it, since
    the algorithm by Har-Peled and Mendel~\cite[Sec.~3.1]{Har-Peled_2006_nets-algorithm} can be adapted to the $k$-distance case in arbitrary metric spaces with little effort.
    Also, the friends of $p_i$ (Section~\ref{sec:algorithm}) can be computed with a slight adaptation of their techniques; we used quadtreaps mostly for the
    ease of presentation. However, the computation of critical values of simplices described in Section~\ref{sec:algorithm} relies
    on computing minimum enclosing balls of balls, for which we used an efficient algorithm for the Euclidean case.
    The complexity of this step remains unspecified for a general metric space. This is common in related work; see, for instance~\cite[Sec.~5]{cavanna-2015-geometric_sparse}.
    
    Another natural goal is to remove the dependence on the spread. This dependence is caused by the computation of the \permut which is inspired by
    the algorithm of~\cite[Sec.~3.1]{Har-Peled_2006_nets-algorithm}. In the same paper~\cite[Sec.~3.2--3.3]{Har-Peled_2006_nets-algorithm},
    they describe an approach to remove the spread from the bound (for $k=1$) using an approximate version of the greedy permutation.
    While our construction of the sparsified filtration can be easily adapted to work with an approximate version of the \permut,
    it seems less straight-forward to generalize their techniques to the $k$-distance, even in the Euclidean case.

    The \permut relates to the \emph{Distance to Measure} (DTM)~\cite{Chazal_Cohen-Steiner_Merigot:11:inference_for_prob}, which is the square average of the distances to the $k$ nearest neighbors. The DTM has the advantage of being 
    robust, in terms of the Wasserstein distance, to perturbations on the sample~\cite[Sec.~3]{Chazal_Cohen-Steiner_Merigot:11:inference_for_prob}. However, most of the existing methods for sparsifying filtrations obtained via the DTM~\cite{buchet-sheehy_2016_persist-homol-measures, Anai_Chazal_etc-2019-DTM_filtrations, guibas_witnessed_k_distance} require a preliminary approximation by weighted distances. 
    Our approach might be adaptable to directly sparsify DTM filtrations.
    
    Our techniques also yield a linear-sized approximation of the multicover bifiltration up to $\maxk$. 
    However, the presence of an exponential factor on $\maxk$ in our size bounds suggests a restriction of our approach's usability to small values of $\maxk$. 
    The exponential factor also carries over to the expected computation time. Reducing that dependency on $\maxk$ is another possible line of future work. 
    Note that~\cite[Prop.~5]{kerber_corbet-2023-multicover_bifiltration} gives a size bound of $\mathcal{O}\left(n^{d+1}\right)$ for the exact version, 
    but we ask whether a polynomial bound on $\maxk$ could be achieved without such a blow-up in the dependency on $n$.

    We point out in this context that the double-nerve construction of \cite{kerber_corbet-2023-multicover_bifiltration} could be applied in Section~\ref{sec:multicovers}:
    instead of taking the nerve over all $\ell$-cones with $k\leq\ell\leq\maxk$ at level $k$, we could as well only take the union of all nerves of the cones at level $\ell$ and $\ell+1$
    with $k\leq \ell\leq \maxk-1$. The proof of equivalence is via a mapping telescope construction, similar to \cite[Thm 4]{kerber_corbet-2023-multicover_bifiltration}.
    While this approach yields a smaller simplicial bifiltration, the complexity bounds do not change, so we omit the details.
    
    Finally, a natural question is the practicality of our algorithm. We remark that even for $k=1$, while some work has been devoted to practical aspects of computing sparsifications~\cite{blaser_brun-2019-sparse-practice,sparips,tamal_shi_wang-2019-simba,gudhi:RipsComplex,burella-2021-giotto_for_rips}, the actual practical computation is still an unresolved problem. We think that the natural order for a practically efficient solution would be to first identify best practices in the simpler $k=1$ case and subsequently try to adapt them to larger values of $k$. So, while we would be curious about the performance of our algorithm, such an evaluation seems to be premature at the moment.    

\bibliographystyle{plain}
\bibliography{references}

\end{document}